\documentclass{jaa}
\usepackage{natbib}
\usepackage{graphicx}
\usepackage{verbatim}

\begin{document}\sloppy

\newcommand{\hi}{\mbox{\rm H\,{\sc i}}}
\newcommand{\hii}{\mbox{\rm H\,{\sc ii}}}
\newcommand{\mgii}{\mbox{\rm Mg\,{\sc ii}}}
\newcommand{\mgi}{\mbox{\rm Mg\,{\sc i}}}
\newcommand{\feii}{\mbox{\rm Fe\,{\sc ii}}}
\newcommand{\mnii}{\mbox{\rm Mn\,{\sc ii}}}
\newcommand{\crii}{\mbox{\rm Cr\,{\sc ii}}}
\newcommand{\tii}{\mbox{\rm Ti\,{\sc ii}}}
\newcommand{\znii}{\mbox{\rm Zn\,{\sc ii}}}
\newcommand{\caii}{\mbox{\rm Ca\,{\sc ii}}}
\newcommand{\nai}{\mbox{\rm Na\,{\sc i}}}
\newcommand{\ci}{\mbox{\rm C\,{\sc i}}}
\newcommand{\cii}{\mbox{\rm C\,{\sc ii}}}
\newcommand{\ciii}{\mbox{\rm C\,{\sc iii}}}
\newcommand{\civ}{\mbox{\rm C\,{\sc iv}}}
\newcommand{\ovi}{\mbox{\rm O\,{\sc vi}}}
\newcommand{\sitwo}{\mbox{\rm Si\,{\sc ii}}}
\newcommand{\sithree}{\mbox{\rm Si\,{\sc iii}}}
\newcommand{\sifour}{\mbox{\rm Si\,{\sc iv}}}
\newcommand{\ha}{\rm H\,$\alpha$}
\newcommand{\hb}{\rm H\,$\beta$}
\newcommand{\oi}{\rm [O\,{\sc i}]}
\newcommand{\oii}{\rm [O\,{\sc ii}]}
\newcommand{\oiii}{\rm [O\,{\sc iii}]}
\newcommand{\nii}{\rm [N\,{\sc ii}]}
\newcommand{\sii}{\rm [S\,{\sc ii}]}
\newcommand{\lya}{\ensuremath{{\rm Ly}\alpha}}
\newcommand{\lyb}{\ensuremath{{\rm Ly}\beta}}
\newcommand{\lymana}{\ensuremath{{\rm Lyman}-\alpha}}
\newcommand{\chisq}{$\chi^2_{\nu}$}
\newcommand{\zabs}{$z_{\rm abs}$}
\newcommand{\zqso}{$z_{\rm QSO}$}
\newcommand{\zem}{$z_{\rm em}$}
\newcommand{\nhi}{$N$($\hi$)}
\newcommand{\wmgii}{$\rm W_{\rm r}^{\mgii}$}
\newcommand{\mstar}{$\rm M_*$}
\newcommand{\fc}{$f_{\rm c}$}
\newcommand{\ts}{$T_{\rm s}$}
\newcommand{\tk}{$T_{\rm k}$}
\newcommand{\taudv}{$\int\tau {\rm dv}$}
\newcommand{\kms}{km\,s$^{-1}$}
\newcommand{\cms}{cm$^{-2}$}
\newcommand{\cc}{cm$^{-3}$}
\newcommand{\mjb}{mJy~beam$^{-1}$}
\newcommand{\ergs}{$\rm erg\,s^{-1}$}
\newcommand{\ergscm}{$\rm erg\,s^{-1}\,cm^{-2}$}
\newcommand{\ergscmarc}{$\rm erg\,s^{-1}\,cm^{-2}\,arcsec^{-2}$}
\newcommand{\msun}{$\rm M_\odot$}
\newcommand{\msunyr}{$\rm M_\odot yr^{-1}$}
%

\title{Probing galaxy evolution through \hi\ 21-cm emission and absorption: current status and prospects with the Square Kilometre Array}


\author{Rajeshwari Dutta\textsuperscript{1,2*}, 
Sushma Kurapati\textsuperscript{3}, 
J. N. H. S. Aditya\textsuperscript{4,5},
Omkar Bait\textsuperscript{6},
Mousumi Das\textsuperscript{7},
Prasun Dutta\textsuperscript{8},
K. Indulekha\textsuperscript{9},
Meera Nandakumar\textsuperscript{8},
Narendra Nath Patra\textsuperscript{10},
Nirupam Roy\textsuperscript{11},
Sambit Roychowdhury\textsuperscript{5,12}
}
\affilOne{\textsuperscript{1}Dipartimento di Fisica G. Occhialini, Universit\`{a} degli Studi di Milano Bicocca, Piazza della Scienza 3, 20126 Milano, Italy\\}
\affilTwo{\textsuperscript{2}INAF - Osservatorio Astronomico di Brera, via Bianchi 46, 23087 Merate (LC), Italy\\}
\affilThree{\textsuperscript{3}Department of Astronomy, University of Cape Town, Private Bag X3, Rondebosch 7701, South Africa\\}
\affilFour{\textsuperscript{4}Sydney Institute for Astronomy, School of Physics A28, The University of Sydney, NSW 2006, Australia\\}
\affilFive{\textsuperscript{5}ARC Centre of Excellence for All Sky Astrophysics in 3 Dimensions (ASTRO 3D), Australia\\}
\affilSix{\textsuperscript{6}Observatoire de Gen\`eve, Universit\'e de Gen\`eve, 51 Ch. des Maillettes, 1290 Versoix, Switzerland\\}
\affilSeven{\textsuperscript{7}Indian Institute of Astrophysics, II Block, Koramangala, 560 034, Bangalore, India\\}
\affilEight{\textsuperscript{8}Department of Physics, IIT (BHU) Varanasi, 221005 India\\}
\affilNine{\textsuperscript{9}School of Pure and Applied Physics, Mahatma Gandhi University, Kottayam, India\\}
\affilTen{\textsuperscript{10}Department of Astronomy, Astrophysics and Space Engineering, Indian Institute of Technology Indore, Indore 453552, India\\}
\affilEleven{\textsuperscript{11}Department of Physics, Indian Institute of Science, Bangalore 560012, India\\}
\affilTwelve{\textsuperscript{12}International Centre for Radio Astronomy Research (ICRAR), University of Western Australia, 35 Stirling Highway, Crawley, WA 6009, Australia\\}


\twocolumn[{

\maketitle

\corres{rajeshwaridutta21@gmail.com}

\msinfo{28 February 2022}{27 May 2022}

\begin{abstract}
One of the major science goals of the Square Kilometre Array (SKA) is to understand the role played by atomic hydrogen (\hi) gas in the evolution of galaxies throughout cosmic time. The hyperfine transition line of the hydrogen atom at 21-cm is one of the best tools to detect and study the properties of \hi\ gas associated with galaxies. In this article, we review our current understanding of \hi\ gas and its relationship with galaxies through observations of the 21-cm line both in emission and absorption. In addition, we provide an overview of the \hi\ science that will be possible with SKA and its pre-cursors and pathfinders, i.e. \hi\ 21-cm emission and absorption studies of galaxies from nearby to high redshifts that will trace various processes governing galaxy evolution.
\end{abstract}


}]


\doinum{12.3456/s78910-011-012-3}
\artcitid{\#\#\#\#}
\volnum{000}
\year{0000}
\pgrange{1--}
\setcounter{page}{1}
\lp{1}

\section{Introduction}
\label{sec:introduction}

Galaxy evolution is driven by the interplay between stars and different phases of gas. Hydrogen, the most abundant element in the Universe, is present in a wide range of gas phases \citep[e.g.][]{mckee1977,kulkarni1988,wolfire1995,snow2006} -- cold molecular (H$_2$) gas (temperature, $T\sim$10\,K; density, $n\gtrsim$10$^3$\,\cc), cold ($T\sim$100\,K; $n\sim$30\,\cc) and warm ($T\sim$10$^4$\,K; $n\sim$0.3\,\cc) atomic (\hi) gas, warm ionized (\hii) gas ($T\sim$10$^4$\,K; $n\sim$0.3\,\cc) and hot ionized gas ($T\sim$10$^6$\,K; $n\sim$10$^{-3}$\,\cc). Hydrogen is also observed in a wide range of structures, from molecular clouds and \hii\ regions in the interstellar medium (ISM) of galaxies, atomic discs and the circumgalactic medium \citep[CGM; typically defined as the region within the virial radius or a few 100\,kpc around galaxies;][]{tumlinson2017}, to the large-scale intergalactic medium (IGM).

Galaxies initially evolved using their hydrogen reservoir obtained during reionization, which was subsequently replenished due to accretion from the IGM and mergers. The ionized gas accreted from the IGM and the CGM converts to atomic hydrogen gas that can eventually convert to molecular gas in the ISM. This involves multiple processes such as gas cooling, radiative recombination and molecule formation on dust grains. The molecular gas phase is the reservoir from which stars are formed in the ISM. The atomic gas thus constitutes a crucial intermediary phase in the cosmic baryon cycle that has a direct impact on the star formation rates, stellar masses and metallicities of the galaxies. While at one hand the atomic gas modulates the star formation activity in galaxies, it also gets affected by the radiative, chemical and mechanical feedback associated with star formation \citep[e.g.][]{cox1974,wolfire1995,gent2013,gatto2015,naab2017}. Therefore, in order to understand the physical processes that drive the cosmic evolution in the global star formation rate density \citep{madau2014}, it is vital to trace the evolution of the atomic gas associated with galaxies. The galaxy ecosystem is additionally regulated by the large-scale environment that is observed to significantly affect the morphology and star formation rate (SFR) of galaxies \citep[e.g.][]{dressler1980,balogh99,baldry2006,peng2010,fossati2017}. The atomic gas in gas-rich galaxy discs is more extended than the stellar component and is therefore highly vulnerable to distortions caused by environmental effects such as tidal interactions and ram pressure stripping \citep[e.g.][]{yun1994,sancisi2008,chung2009,mihos2012,brown2017}. Hence, the hierarchical structure formation process of galaxies including merger and accretion leave discernible imprints in the atomic gas phase. 

The \hi\ 21-cm line is one of the best probes of the atomic hydrogen gas in and around galaxies. The hydrogen atom emits a photon at 21-cm or 1420 MHz during the transition of an electron between the hyperfine levels in the $1s$ ground state. Despite being a highly forbidden transition, the \hi\ 21-cm line is observable thanks to the large amount of atomic hydrogen in the Universe. Occurring in radio frequencies, this radiation from the hydrogen atom penetrates through dust clouds and provides us a more complete view of the ISM than that by visible light. H. C. van de Hulst first predicted the observability of the \hi\ 21-cm line in 1944. Subsequently, \hi\ 21-cm emission from the Milky Way was observed by \citet{ewen1951}, \citet{muller1951} and \citet{pawsey1951}. When it comes to \hi\ 21-cm absorption, the first detections of absorption from the Milky Way were reported by \citet{hagen1954a} and \citet{hagen1954b}. Thereafter, \citet{roberts1970} detected for the first time \hi\ 21-cm absorption associated with an extragalactic radio source, Centaurus A, followed by the observation of intervening \hi\ 21-cm absorption at $z=0.69$ towards the background radio source 3C~286 by \citet{brown1973}.

Since the first observations, the \hi\ 21-cm line has proved to be a powerful probe of the structure, kinematics and physical conditions of the atomic gas in our own and other galaxies. One of the major science drivers behind the creation of the Square Kilometre Array (SKA) is to understand the role of \hi\ gas in the formation and evolution of galaxies \citep{staveley-smith2015}. The objective of this review article is to provide an overview of the various \hi\ 21-cm studies to date of galaxies from nearby to high redshifts, both via emission (Section~\ref{sec:emission}) and absorption (Section~\ref{sec:absorption}), and the \hi\ galaxy science that is being/will be carried out with SKA and its pre-cursors and pathfinders (Section~\ref{sec:ska}). For cosmological \hi\ 21-cm studies, we refer the readers to other reviews in the same volume that cover aspects related to the Epoch of Reionization (EoR), post-EoR \hi\ power spectra and \hi\ intensity mapping.

\section{\hi\ 21-cm emission}
\label{sec:emission} 

Over the last two decades, much progress has been made in observations of \hi\ gas in the Milky Way (MW) and nearby galaxies. Several Galactic all-sky low-resolution surveys such as the Effelsberg-Bonn \hi\ Survey \citep[EBHIS;][]{kerp11}, the Galactic All-Sky Survey \citep[GASS;][]{mcclure09}, the Galactic Arecibo L-band Feed Array \hi\ survey \citep[GALFA-HI;][]{peek11}, and the \hi\ 4$\pi$ survey \citep[HI4PI;][]{bekhti16} were used to unravel the volume density distribution of the gaseous disk up to its borders, to study the MW halo \citep{ford08, winkel11, moss13, hammer15, lenz16, kerp16} and the disk-halo interaction \citep{ford10, mcclure10, lenz15, rohser16}. The high resolution Galactic plane surveys such as the Canadian Galactic Plane Survey \citep[CGPS;][]{taylor03}, the Southern Galactic Plane Survey \citep[SGPS;][]{mcclure05}, the VLA Galactic Plane Survey \citep[VGPS;][]{stil06}, and the \hi, OH, Recombination line survey of the Milky Way \citep[THOR;][]{beuther16} have been instrumental in the study of the multiphase structure of \hi\ gas \citep[e.g.][]{strasser07, dickey09} and in the discovery of Galactic shells and filaments that give clear evidence of the injection of energy into the ISM by supernova explosions \citep[e.g.][]{dawson11}. A detailed discussion of \hi\ studies of the MW, however, is beyond the scope of this article.

In the following sections, we discuss how \hi\ 21-cm emission observations allow us to understand various aspects of nearby galaxies such as the relationship between star formation and gas (Section~\ref{sec:emission_starformation}), the effect of environment on galaxies (Section~\ref{sec:emission_environment}), the distribution of dark matter and angular momentum (Section~\ref{sec:emission_kinematic}), structure of \hi\ disk (Section~\ref{sec:emission_vertical}), turbulence in the ISM (Section~\ref{sec:emission_turbulence}), the relationship between \hi\ gas and galaxy properties (Section~\ref{sec:emission_scaling}), and the average neutral gas content of galaxies (Section~\ref{sec:emission_stacking}). 

\subsection{\hi\ emission surveys}
\label{sec:emission_surveys}

Several \hi\ emission line surveys on single dish radio telescopes such as the \hi\ Parkes All Sky Survey \citep[HIPASS;][]{meyer04} and the Arecibo Fast Legacy ALFA survey \citep[ALFALFA;][]{giovanelli05} have been used to measure the global \hi\ properties of nearby galaxies such as the \hi\ mass function \citep[HIMF; the distribution function of galaxies as a function of \hi\ mass;][]{zwaan03, zwaan2005, martin10, haynes11}, contribution of different galaxy populations to the HIMF \citep{dutta20, dutta21}, environmental dependence of the HIMF \citep{moorman14, jones18, said19}, the clustering of \hi-selected galaxies \citep{passmoor11, martin12, papastergis13, guo17}, the \hi\ velocity width function \citep{moorman14}, and scaling relations of atomic gas fraction with galaxy parameters \citep[e.g.][]{catinella2010,catinella13}. 

At the same time, spatially-resolved observations of nearby galaxies are critical to understand the structure and dynamics of galaxies. A number of \hi\ emission line surveys on radio interferometers such as ``The Westerbork \hi\ Survey of Irregular and Spiral Galaxies'' \citep[WHISP;][]{vanderhulst2001}, ``The \hi\ Nearby Galaxy Survey'' \citep[THINGS;][]{walter2008}, ``Faint Irregular Galaxies GMRT Survey'' \citep[FIGGS;][]{begum2008a}, ``The Local Volume \hi\ Survey'' \citep[LVHIS;][]{koribalski2008}, ``Survey of \hi\ in Extremely Low-mass Dwarfs'' \citep[SHIELD;][]{cannon2011}, ``The Westerbork Hydrogen Accretion in Local Galaxies survey'' \citep[HALOGAS;][]{heald2011}, ``Local Irregulars That Trace Luminosity Extremes, The \hi\ Nearby Galaxy Survey'' \citep[LITTLE THINGS;][]{hunter2012} and ``Very Large Array survey of ACS Nearby Galaxy Survey Treasury galaxies'' \citep[VLA-ANGST;][]{ott2012} have increased the number of galaxies for which high-quality, high-resolution \hi\ emission data are available. For example, THINGS, a high spectral ($\le5$\,\kms) and spatial ($\approx6$'') resolution \hi\ emission survey using the Very Large Array (VLA), has observed 34 nearby galaxies at distances 3$<$D$<$15\,Mpc leading to a spatial resolution of $\approx100-500$\,pc. While the THINGS sample comprises of galaxies with \hi\ masses of $(0.01-14)\times10^9$\,\msun, LITTLE THINGS focuses on nearby dwarf galaxies using VLA \hi\ observations of 37 dwarf irregular and 4 blue compact dwarf galaxies at D$<$10\,Mpc. FIGGS, along with its extension FIGGS2 \citep{patra2016}, is a much larger survey of \hi\ gas in faint dwarf irregular galaxies, comprising of $\sim$75 galaxies (median \hi\ mass $3\times10^7$\,\msun) at D$<$10\,Mpc, observed with high spectral resolution ($\le $3\,\kms) and multi-spatial resolution (5'' to 40'') observations using the Giant Metrewave Radio Telescope (GMRT). Observations from the above surveys have been used to investigate the relation between \hi\ gas and star formation, galaxy morphology, mass and environment, as discussed in the following sections.

\subsection{Star formation in galaxies}
\label{sec:emission_starformation}

The process of star formation is central to galaxy evolution since it drives the consumption of gas in galaxies as well as the evolution of gas in the baryon cycle. Here we discuss star formation and its relation with \hi\ gas.

\subsubsection{Gas accretion:} 
\label{sec:emission_accretion}

Galaxies accrete gas from the IGM in order to sustain their star formation via two modes \citep{keres05, dekel09a}. In the first mode called as the `hot mode' (T $\sim$ 10$^{\rm 6}$\,K) accretion, the gas is shock heated once it enters the dark matter halo and remains part of the hot halo. The gas will cool radiatively if it reaches sufficiently high density and settle onto a disc \citep{rees77}. Recent simulations have highlighted the importance of a second mode of accretion called `cold mode' (T $\sim$ 10$^{\rm 4}$\,K) accretion, where cold gas flows along filaments directly into the disk. `Cold mode' accretion is predominant in low-mass (halo mass M$_{\rm h} \lesssim10^{11}$\,\msun) galaxies residing in low-density and high-redshift environments, while `hot mode' accretion dominates in high-mass (M$_{\rm h} \gtrsim10^{11}$\,\msun) galaxies evolving in high-density environments. While direct kinematic evidence of gas accretion is rare, there are potential indirect signatures of gas accretion onto galaxies \citep{putman2017}, including observations of \hi\ filaments and tails around galaxies, extra-planar \hi\ gas clouds and extended, warped or lopsided \hi\ spiral discs \citep[e.g.][]{sancisi2008,deblok2014,leisman2016,xu2021}. 

\subsubsection{Star formation, stellar mass and metallicity:} 
\label{sec:emission_stellarmass}

The scaling of star formation rate (SFR) with other galaxy properties such as the mass (or surface densities) of stars and cold gas provides insights into the complex processes governing galaxy evolution such as efficiency of gas consumption, mechanisms for gas replenishment and connection between the instantaneous SFR and gas reservoir. The SFR and stellar masses (M$_{\rm \ast}$) of star-forming galaxies are observed to correlate with each other, which is termed as the `main sequence' \citep{noeske07, elbaz07}. Galaxies falling on the main sequence are typically young, actively star-forming blue disks. The more massive, passive galaxies form the `red sequence' which comprises of old stellar populations where star formation has been quenched. This quenching could arise as a result of gas depletion due to environmental interactions as discussed in Section~\ref{sec:emission_environment}. However, few quenched galaxies are found to have large reservoirs of \hi\ gas despite very little star formation \citep[e.g.][]{Gereb16, Gereb18, Bait20}. The \hi\ gas distribution in these galaxies has diverse nature such as large and diffuse rings, low surface brightness discs and counter rotation between the gas and the stars. The gas depletion time-scales in these galaxies are extremely high ($\sim$10-100 Gyrs) owing to their large gas fractions and low SFRs. 

Along with SFR and M$_{\rm \ast}$, the gas phase metallicities ($Z$) in the ISM represent the fundamental quantities of galaxies. The more massive galaxies are found to be more chemically enriched \citep[the mass-metallicity relation, MZR;][]{lequeux1979}, while galaxies with higher SFRs tend to have lower metallicities for a given stellar mass \citep[the fundamental metallicity relation, FMR;][]{mannucci2010}. It has been suggested that the FMR could be driven by a more fundamental relation between stellar mass, metallicity and \hi\ mass or neutral gas fraction \citep{bothwell2013,lagos2016,derossi2017}. The dependence of metallicity on SFR could be a byproduct of the dependence on the gas density, via the Kennicutt-Schmidt relation that we discuss next.

\subsubsection{Star formation and \hi\ gas:} 
\label{sec:emission_kennicutt}

\begin{figure}[!t]
\centering
\includegraphics[width=0.45\textwidth]{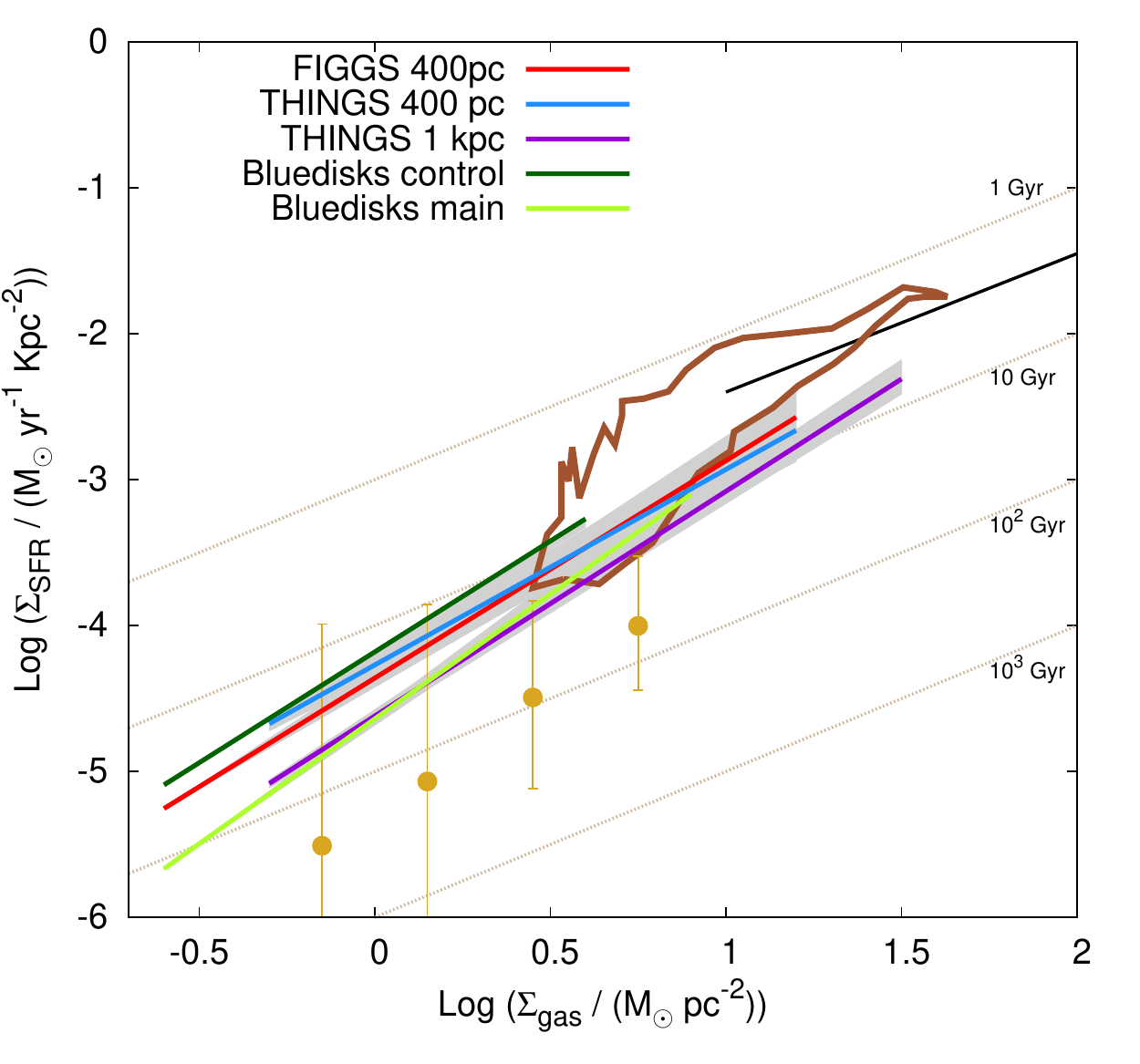}
\caption{
The resolved mean Kennicutt-Schmidt (KS) relation using $\Sigma_{\rm HI}$ for $\Sigma_{\rm gas}$ from \citet{roychowdhury15}, shown with the labelled coloured lines for dwarf irregular galaxies from the FIGGS survey \citep{begum2008a}, and \hi\ dominated regions of spirals galaxies from the THINGS \citep{walter2008} and Bluedisk \citep[][$\sim$10 kpc scales]{wang13} surveys. The 1$\sigma$ error on the fits and their overlap are represented by the grey shaded area. Dotted beige lines indicate various constant gas depletion timescales. The black line represents the $\Sigma_{\rm H_2}$ based KS law for nearby spirals from \citet{leroy13}, the brown contour indicates the relation as seen on 750\,pc scales within the optical discs of nearby spirals by \citet{bigiel08}, and the yellow open circles represent the median $\Sigma_{\rm SFR}$ in similar $\Sigma_{\rm gas}$ bins in the very outer discs of spirals from the THINGS survey measured by \citet{bigiel2010} at 1\,kpc scales with the scatter indicated by the error bars. In contrast to \citet{bigiel2010}, \citet{roychowdhury15} measured the $\Sigma_{\rm SFR}$ more accurately to arrive at the mean KS relations while foregoing the sensitivity to the scatter in the relations.}
\label{fig_kslaw}
\end{figure}

The surface density of SFR ($\Sigma_{\rm SFR}$) is observed to strongly correlate with the cold gas surface density ($\Sigma_{\rm gas}$) in late-type galaxies, which is called as the Kennicutt–Schmidt (KS) law \citep{schmidt59, schmidt63, kennicutt89, kennicutt98}:

 \begin{equation}
     \Sigma_{\rm SFR} = {\rm A} \Sigma_{\rm gas}^{\rm N}
     \label{eqn_kennicutt}
 \end{equation}
 
\noindent 
where $\Sigma_{\rm SFR}$ is the surface density of SFR (in M$_{\odot}$ kpc$^{-2}$ yr$^{-1}$), $\Sigma_{\rm gas}$ is the gas surface density (in M$_{\odot}$ pc$^{-2}$), and A is a constant of proportionality. \citet{kennicutt98} found index N$\sim$1.4 using the sum of \hi\ and molecular gas (H$_{2}$) for the total gas surface density and for values averaged over galactic disks. However, varying values of N were found in different studies, especially with spatially resolved measurements of the KS law \citep{bigiel08}. \citet{delosreyes19} revisited the KS law in local star-forming and dwarf galaxies on galaxy-averaged scales, and found that spirals lie on a tight relation (with N = 1.41 $\pm$ 0.07) while dwarfs lie below this relation. 

When the KS law was explored by focusing only on the molecular phase of the gas, i.e. the phase more directly related to star formation, spatially resolved studies generally found N$\sim$1 \citep[e.g.][but see also \citet{shetty2014}]{leroy13}. At the same time, observational studies always found a KS law with N$\sim$1.4 when only the atomic gas in galaxies was considered, albeit inefficient compared to the molecular phase. This was especially so for star-forming dwarf galaxies and low surface brightness (LSB) galaxies whose ISM is atomic gas dominated, no matter whether the KS law used values averaged over galaxy disks or used spatially resolved values \citep{wyder09,roychowdhury2009,roychowdhury2011,roychowdhury2014,bigiel2010,elmegreen2015,patra16}. \citet{bigiel2010} claimed that in the outer regions of galaxy disks of normal galaxies, the spatially resolved KS relation has a steep slope. \citet{roychowdhury15} though looked at the spatially resolved KS law in \hi-dominated regions of both normal and dwarf galaxies including the outer regions of normal galaxies (Fig.~\ref{fig_kslaw}), and found a KS law with only a slightly steeper slope of N$\sim$1.5. This was found to be a factor of 10 or more inefficient compared to the molecular gas KS law. Through comparison with simulations, they suggested that the atomic gas KS law exists because stellar and supernova feedback set up the physical conditions in ISM that is dominated by the atomic phase of gas.

\subsection{Galaxy environment}
\label{sec:emission_environment}

The distribution of galaxies in the Universe forms a cosmic web, consisting of voids, filaments, and clusters. Here we discuss properties and \hi\ observations of galaxies residing in these different environments.

\subsubsection{Clusters and groups:}
\label{sec:emission_clusters}

There is well-established observational evidence that galaxy properties such as SFR \citep[e.g.][]{balogh99, poggianti06}, morphology \citep[e.g.][]{dressler1980}, and gas content \citep[e.g.][]{denes14} have a strong dependence on the environment. \citet{dressler1980} found that red, gas-poor, elliptical (early-type) galaxies are more common in high-density environments such as clusters, whereas blue, gas-rich, spiral (late-type) galaxies dominate the field population (known as the `density-morphology relation'). Similarly, the fraction of blue star-forming spiral galaxies in clusters is higher at intermediate reshifts than that in local clusters \citep[termed as the `Butcher-Oemler effect';][]{butcher78}. These effects suggest that galaxies in high-density environments lose their gas quickly and stop forming stars, while galaxies in low-density environments tend to accrete fresh gas and keep forming stars. 
Various possible processes have been proposed for the transformation of star-forming spiral galaxies into passive elliptical galaxies in clusters. These include: (i) strong ram pressure stripping \citep[e.g.][]{gunn72} that can strip the cold gas that fuels star formation, leading to quenching on short time scales \citep[e.g.][]{roediger06, bekki14, boselli14}; (ii) galaxy strangulation/starvation, where the weakly-bound diffuse hot halo gas from the galaxy is stripped leading to quenching of the star formation once the gas supply is exhausted \citep[e.g.][]{larson80}; (iii) tidal interactions that lead to distortions of stellar and gas distributions \citep[e.g.][]{byrd90, moore98}.

The \hi\ disks of galaxies are ideal tracers of such environmental processes due to their sensitivity to external perturbations. Studies of statistically large samples show that galaxies in high-density environments tend to have lower \hi\ gas than those in average-density environments \citep[e.g.][]{giovanelli85, solanes01, chung2009, healy21}. The degree of \hi\ depletion in groups and clusters is found to be related to the morphology of galaxies, with early-type and dwarf spirals being more depleted \citep{solanes01, kilborn09}. The high-density environment is also found to affect the connection of \hi\ gas with galaxy properties, with the scaling relations of \hi\ with stellar mass and colour of galaxies being significantly offset towards lower gas fractions in clusters \citep{cortese2011}. 

High-resolution \hi\ observations have revealed signatures of ram pressure stripping such as truncated \hi\ discs, one-sided \hi\ tails, lopsided \hi\ morphologies and enhanced SFR on the leading side for galaxies in nearby clusters \citep[e.g.][]{cayatte90, vollmer01, chung2009, yoon17}, and also in groups \citep[e.g.][]{verdes01, rasmussen08, jaffe12, hess13}. Further, \hi\ observations of galaxies hosting transient phenomena such as Gamma Ray Bursts, supernovae, Fast-Evolving Luminous Transients and Fast Radio Bursts have found disturbed \hi\ gas and high column density rings indicative of mergers and interactions with nearby galaxies in dense environment \citep{arabsalmani2015,arabsalmani2019,roychowdhury2019,kaur2022}. Studies of clusters and groups have also provided evidence that galaxies in cluster outskirts and groups undergo `pre-processing' before they fall into the cluster itself \citep[e.g.][]{zabludoff98, poggianti99, kantharia05, catinella13, hess13}. Particularly for satellite galaxies, the environmental suppression of \hi\ gas at fixed \mstar\ and specific SFR is shown to begin in the group regime, before galaxies reach the cluster environment \citep{brown2017}. A recent study has revealed that there are large quantities of unaccounted for \hi\ gas in the intergalactic space between galaxies of low mass (\mstar\ $\lesssim 10^{9.5}$\,\msun) groups \citep{roychowdhury2022}, which might be a pointer towards `pre-processing' of \hi\ gas during group assembly.

\subsubsection{Voids:} 
\label{sec:emission_voids}

In the other extreme, voids provide us with environments that are largely unaffected by the complex processes in high-density environments. Void galaxies are expected to evolve in relative solitude and be less evolved since they typically form at later stages than those in denser regions \citep[e.g.][]{aragon13}. Observations show that void galaxies are in general smaller, bluer, later type, with higher SFRs than their counterparts in the field \citep[e.g.][]{rojas04, patiri06, von08,  hoyle12, moorman14, moorman16}. However, it was suggested that the difference in SFR of void galaxies could be attributed to the `morphology-density relation' \citep[see][]{patiri06, park07}. Indeed, at a fixed luminosity and morphology, the brighter void galaxies (M$_{\rm r} < -16.5$) have statistically identical colours and SFRs as those of galaxies in the field \citep[e.g.][]{kreckel12}, while the fainter dwarfs in voids have higher specific SFRs than their average density counterparts \citep{von08, moorman16}. 

Similarly, \hi\ gas content of brighter void galaxies was found to be statistically indistinguishable from galaxies in average densities \citep{kreckel12}, while low-luminosity galaxies in voids are systematically more gas-rich than those in denser regions \citep{pustilnik16a}. Recent high-resolution \hi\ surveys with VLA and GMRT have lead to the discovery of several unusual objects that show signs of interaction and gas accretion. For example, approximately linear triplets of gas-rich galaxies \citep[e.g.][]{beygu13, chengalur13, chengalur17} were found, which could be due to material flow along the filament. Other triplets include DDO68, a very metal-poor galaxy, in which two massive components have already merged \citep{ekta08} and the third component is connected to the primary galaxy by a low surface brightness \hi\ bridge \citep{cannon14}. \citet{kreckel11a} found a void dwarf galaxy with an extremely extended \hi\ disk, signs of an \hi\ cloud with anomalous velocity, misalignment between kinematic minor and major axes, and misalignment between \hi\ and optical axes. \citet{kurapati20b} found that \hi\ disks in void galaxies are $\approx4.5-5$ times more extended than that of the optical disks. Thus, recent \hi\ studies provide support for the idea that voids host at least some galaxies with highly unusual evolutionary histories. Further, \citet{kreckel12} observed 55 brighter dwarfs in voids using the Westerbork Synthesis Radio Telescope (WSRT). They detected \hi\ emission in 41 galaxies and identified 18 \hi-rich companions. In their \hi\ observations of 25 gas-rich dwarfs in the Lynx-Cancer void, \citet{kurapati20b} detected 3 galaxies that are part of triplets, 2 that are merger remnants and 6 that have a non-interacting companion. These studies suggest that the small-scale clustering of galaxies in voids could be similar to that in higher density regions. Finally, there have also been detections of molecular gas in the larger void galaxies \citep[e.g.][]{das2015}.

\subsection{Kinematic studies}
\label{sec:emission_kinematic}

The rotation curves of galaxies are derived by measuring velocities of gas or stars at different radii using emission lines in the optical spectra arising from the ionized gas or the \hi\ 21 cm line. The rotation curves derived from \hi\ extend up to large radii compared to the optical rotation curves, thus allowing us to probe the mass distribution out to much larger radii. These are essential for studies of angular momentum and dark matter properties of galaxies as discussed here.

\subsubsection{Tilted ring model:} 
\label{sec:emission_tiltedring}

\begin{figure}[!t]
\centering
\includegraphics[width=0.35\textwidth]{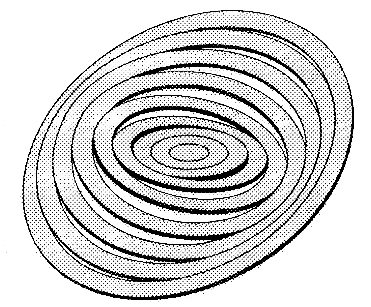}
\caption{Schematic diagram showing the tilted ring model, reproduced from \citet{rogstad74}.}
\label{fig_trm}
\end{figure}

\begin{figure}[!t]
\centering
\includegraphics[width=0.45\textwidth]{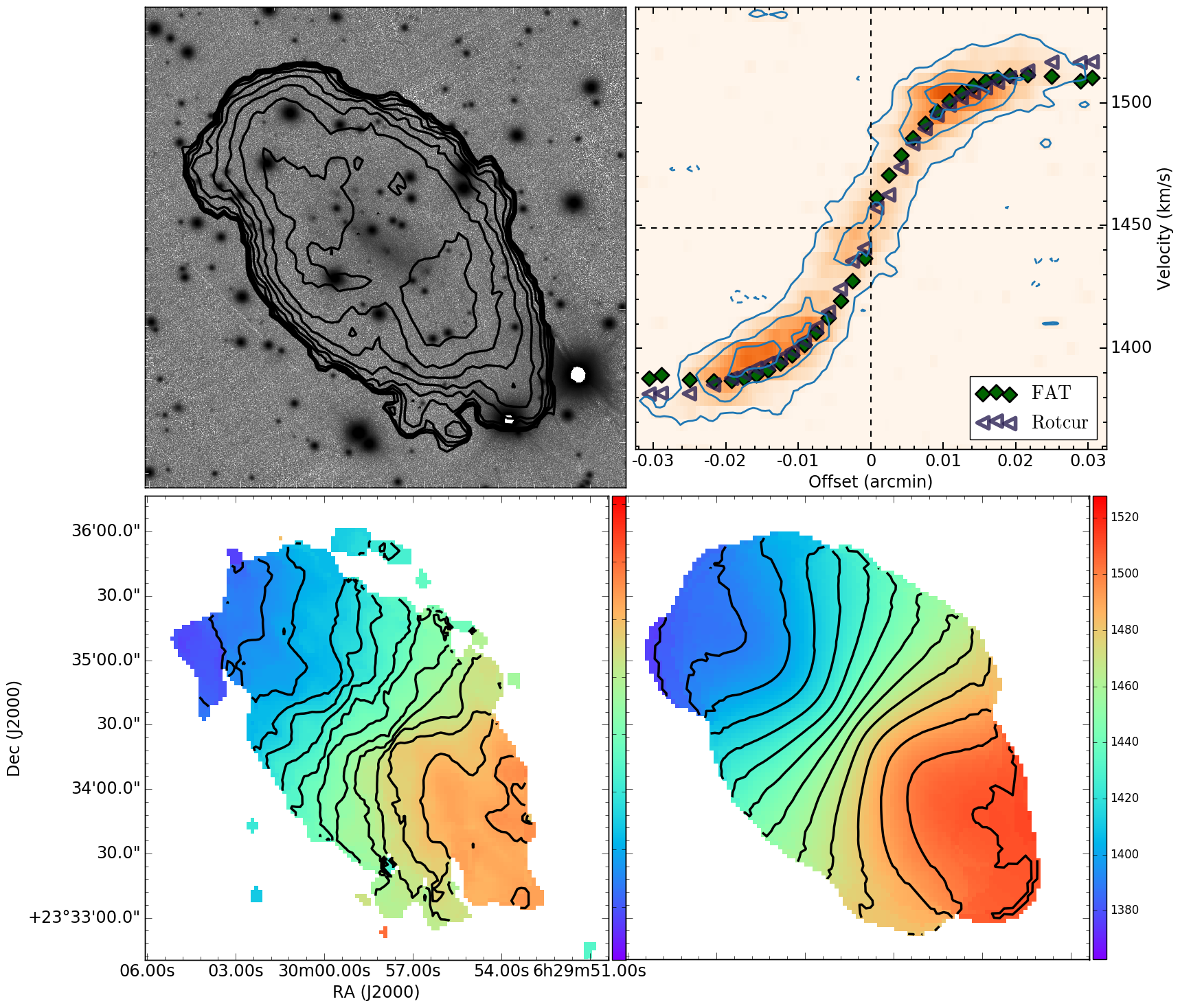}
\caption{\hi\ data and kinematics for the galaxy J0630+23 (from the study of \citet{kurapati20}). Top left: \hi\ distribution of the galaxy overlaid on the SDSS g–band data. Top right: position velocity diagram with the rotation curves obtained using 3D (FAT) and 2D (Rotcur) approaches overlaid on them. Bottom left: velocity field of the data. Bottom right: velocity field of the best fitting FAT model.}
\label{fig_mom_maps}
\end{figure}

The \hi\ kinematics of galaxies have traditionally been modelled using the tilted ring model \citep[Fig.~\ref{fig_trm};][]{rogstad74}, where we assume that the gas is confined to a thin disc and flows at a constant speed at a specific radius from the centre. A rotating disk galaxy can be described by a set of concentric rings, each of which is assumed to be in circular motion and is characterized by centre, systematic velocity, rotation velocity, position angle of the major axis and inclination angle. However, parameters such as centre and position angle may not change much with the radius for well-behaved galaxies. One of the widely used approaches to derive \hi\ rotation curves is by fitting a tilted ring model on the 2D velocity fields that are determined from the 3D data cubes. However, rotation curves derived in this way are expected to suffer from systematic effects such as `beam smearing' (flattening of the velocity gradient in the central region of a galaxy due to finite resolution of the beam). There are software packages available (e.g. TIRIFIC, \citet{jozsa07}; 3D-BAROLO, \citet{diteodoro15}; FAT, \citet{kamphuis15}) that fit the tilted ring model directly on the 3D data cube and include effects such as beam smearing. \citet{kurapati18} find that the rotation curves derived using 3D approach give steeper rotation curves in the inner regions compared to that derived using 2D approach as expected (Fig.~\ref{fig_mom_maps}), since 3D routines are less affected by `beam smearing'.

\subsubsection{Dark matter distribution:}
\label{sec:emission_darkmatter}

It is well known that there is a discrepancy between the amount of mass inferred from the flat rotation curves of spiral galaxies and the visible mass in the form of stars and gas. The commonly used hypothesis to explain this discrepancy is that a halo of unseen dark matter that interacts with the baryonic matter only gravitationally. The dark matter density distribution can be modelled by decomposing the observed rotation curves into contribution from gas, stars and dark matter \citep[e.g.][]{deBlok01, begum04, oh11, oh15, kurapati18b, kurapati20}. The standard $\Lambda$ Cold Dark Matter ($\Lambda$CDM) simulations predict that the dark matter halo has a cuspy density profile, with the density distribution in the inner regions following a power law: $\rho$ $\sim$ r$^{\alpha}$, with $\alpha$ = -1 \citep{navarro96, navarro97}. More recent simulations have shown that the inner slope has a dependence on mass with $\alpha$ lying in the range -0.8 to -1.4 \citep{ricotti03, ricotti07, delpopolo10, delpopolo12, dicintio14}. In contrast to this, observations of dwarf and low surface brightness galaxies, where dark matter dominates throughout the galaxy and the effects of uncertainties in stellar mass to light ratio are minimal, typically find that the dark matter halo has a core towards the centre \citep[e.g.][]{deBlok01, deBlok02, oh11, oh15}. This is called as the `cusp-core problem'.

\begin{figure*}[!t]
\centering
\includegraphics[width=0.45\textwidth]{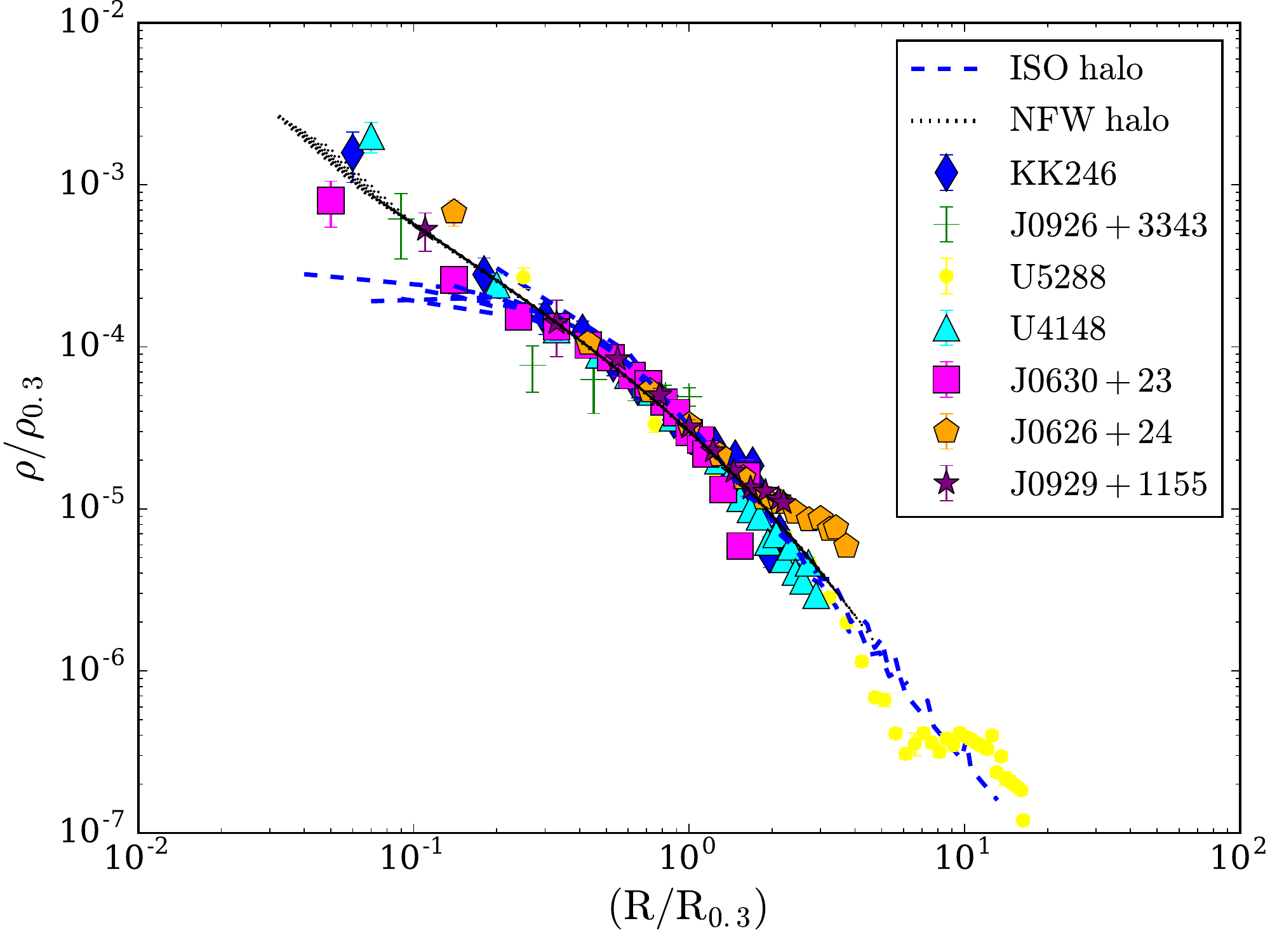}
\includegraphics[width=0.45\textwidth]{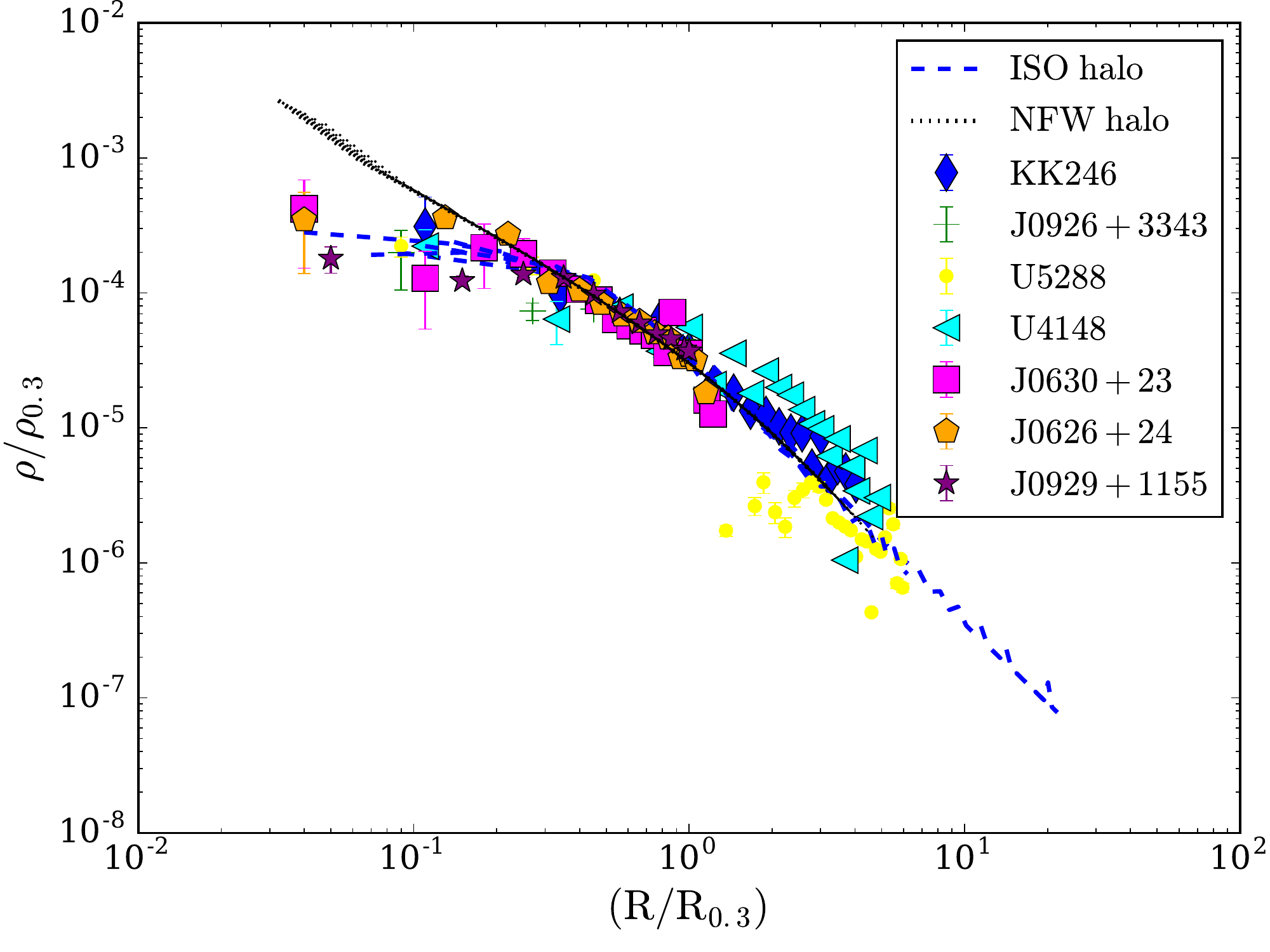}
\caption{The scaled dark matter density profiles (different galaxies are shown with different symbols) derived using 3D approach rotation curves (left) and 2D approach rotation curves (right). The black dotted line represents the NFW halo (with $\alpha$ $\sim$ -1) and the blue dashed line is the best-fit isothermal halo model. Figure reproduced from the study of \citet{kurapati20}.}
\label{fig_dm}
\end{figure*}

Various solutions have been proposed to resolve this problem. For example, baryonic feedback processes were invoked in simulations to generate cores from originally cuspy distributions \citep[see e.g.][]{governato10, pontzen12, read16}. However, there is no complete consensus on the effectiveness of this mechanism since some simulations find that the density profile of dark matter is consistent with cuspy profiles even after including baryonic outflows \citep[e.g.][]{ceverino09, marianacci14}. Other solutions suggest that there are residual systematic effects in determining the rotation curves such as `beam smearing' \citep[e.g.][]{vandenbosch00}, incorrectly measured inclination angles \citep[e.g.][]{rhee04, read16b}, improperly modelled pressure support \citep[e.g.][]{rhee04, valenzuela07, pineda17} or unmodelled non-circular motions \citep[e.g.][]{rhee04, valenzuela07, Oman17}. All of these can lower the inner rotation velocities and thus give a false impression of cores. \citet{kurapati20} investigated the impact of systematic effects on dark matter density profiles by measuring the slopes from the rotation curves derived through 2D and 3D approaches. They found that the average slope obtained from 3D fitting is consistent with the NFW profile, while the slope obtained using the 2D approach is closer to what would be expected for an isothermal profile. Fig.~\ref{fig_dm} shows the plot of the scaled dark matter density versus the scaled radius from \citet{kurapati20}. The left panel shows the density profile obtained using 3D rotation curve, which is consistent with the cuspy NFW profile. The right panel shows the density profile derived using the 2D rotation curve. This is not consistent with the NFW profile but is in good agreement with the best-fit isothermal model in contrast to the 3D rotation curve.
More recently, \citet{ianjamasimanana21} used MeerKAT observations to model the dark matter distribution of a nearby galaxy using a 3D approach. They found that both the NFW and the isothermal models fit the derived rotation curves within the formal errors.

\subsubsection{Angular momentum and mass correlation:}
\label{sec:emission_angular}

The specific angular momentum (j) and mass (M) of galaxies are observed to strongly correlate with each other and with other parameters such as morphology \citep[e.g.][]{fall80, fall83}. In the standard CDM scenario of galaxy formation, galaxies are thought to have acquired their angular momentum through tidal torquing between neighbouring halos \citep{peebles69}. These models predict that j$_{\rm H}$ of dark matter halo scales with its mass as j$_{\rm H} \propto$ M$_{\rm H}^{{2/3}}$. A similar scaling relation is expected for the baryons if one assumes that baryonic matter and dark matter have similar angular momentum distributions at early times and if j$_{\rm b}$ is conserved throughout the formation of galaxies. Indeed, such a scaling relation was obtained for the stellar component, although the constant of proportionality depends on the galaxy morphology \citep{fall83, romanowsky12, fall18}.

However, recent observational studies have indicated that the baryonic relation is more nuanced than a simple j$_{\rm b}$ $\propto$ M$_{\rm b}^{2/3}$ relation. The stellar j$_{\ast}$ -- M$_{\ast}$ relation and gaseous j$_{\rm g}$ -- M$_{\rm g}$ relations were found to follow unbroken power laws, j $\propto$ M$^{\alpha}$, over a wide range of stellar and gas masses, with $\alpha_{\ast}$ $\sim$ 0.55 $\pm$ 0.02 for the j$_{\ast}$ -- M$_{\ast}$ relation for galaxies in the stellar mass range of 10$^{7}$ -- 10$^{11.5}$\,\msun\ \citep{posti18, posti19}, while a steeper slope ($\alpha_{g}$ $\sim$ 1.01 $\pm$ 0.04, \citet{mancera21}; 0.89 $\pm$ 0.05, \citet{kurapati21}) was found for the j$_{\rm g}$ -- M$_{\rm g}$ relation. There has not been a consensus on the nature of the baryonic j$_{\rm b}$ -- M$_{\rm b}$ relation among different authors. \citet{obreschkow14} used all major baryonic components of spiral galaxies and found a tight correlation between j$_{\rm b}$, M$_{\rm b}$ and bulge fraction ($\beta$). They found that the j$_{\rm b}$ $\propto$ M$_{\rm b}^{2/3}$ relation holds across the entire sample, although the galaxies with equal bulge fractions followed $\alpha_{\rm b} \sim $ 1 trend. Some studies have found that the dwarf galaxies have higher j$_{\rm b}$ compared to the bulge-less spirals \citep[e.g.][]{butler17, chowdhury17, kurapati18, kurapati21}, while others have found that the j$_{\rm b}$ -- M$_{\rm b}$ relation is best fit by a single power law \citep[e.g.][]{elson17, lutz18, murugeshan20, mancera21}.

The j$_{\rm b}$ of a galaxy can evolve since various internal and external processes (e.g. accretion, mergers, ram pressure stripping) can redistribute the angular momentum of baryons. \citet{kurapati21} find that the early–type galaxies with large well-rotating disks that show signs of recent gas accretion have excess j$_{\rm g}$ compared to other galaxies. This suggests that the kinematics of the baryonic component of gas-rich galaxies are affected by cold flows and late-stage mergers. In addition, j$_{\rm b}$ may depend on the environment since different processes dominate in different environments. \citet{kurapati18} find that j$_{\rm b}$ of void dwarfs is similar to that of field dwarfs. \citet{murugeshan20} study the effects of the environment on galaxies (in low- and intermediate-density environments) in the atomic gas fraction (f$_{\rm atm}$) -- integrated disc stability parameter (q $\propto$ j$_{\rm b}$ / M$_{\rm b}$) space. They find that galaxies that are currently interacting (or have recently interacted, thereby accreting \hi\ from their smaller companions) show enhanced f$_{\rm atm }$ and lower q values. Using high-resolution ASKAP observations of galaxies in the Eridanus super group, \citet{murugeshan2021} find that these galaxies deviate significantly and lie below the f$_{\rm atm}$--q relation, in contrast to the galaxies in low-density environments that follow the relation consistently.

\subsection{Vertical structure of \hi\ disks}
\label{sec:emission_vertical}

The vertical structure of \hi\ disks is linked with star formation since the conversion of gas into stars critically depends on gas volume densities \citep{bacchini19a}. It further determines the effectiveness of stellar feedback in creating a super-shell or chimney and polluting the CGM with higher metallicity gas. Direct observation of the 3D gas distribution in galaxies is not possible due to the line-of-sight integration effect. However, it is possible to estimate it using theoretical models assisted by observations. A galactic disk is considered as a multi-component system consisting of stellar, atomic, and molecular gas disks in vertical hydrostatic equilibrium under their mutual gravity in the external force field of the dark matter halo. It is then possible to set up the joint Poisson-Boltzmann equation of hydrostatic equilibrium and solve it numerically. The solutions to this equation at every radius can provide a detailed 3D distribution of different galactic components \citep{narayan02a,banerjee08,banerjee10,patra14}.

Using this method, \citet{banerjee11} estimated the \hi\ vertical scale heights in four dwarf galaxies from the FIGGS survey. They found thick \hi\ disks that flare as a function of radius in all their galaxies. A constant vertical gas velocity dispersion (8\,\kms) as a function of radius was used in this study. However, observations show that this velocity dispersion can significantly vary from galaxy to galaxy and as a function of radius. To address this issue, \citet{patra20b} carried out similar modeling in a sample of 23 dwarf galaxies from the LITTLE THINGS survey considering a variable velocity dispersion determined from fitting the \hi\ spectra in different radial annuli. They found that the \hi\ scale height in dwarf galaxies increases from a few hundred parsecs at the center to a few kilo-parsecs at the edge. They also found the median axial ratio to be 0.4 that is much higher than that in spiral galaxies. These studies conclude that thick \hi\ disks in dwarf galaxies originate naturally under the assumption of hydrostatic equilibrium. \citet{patra20c} further used the same technique on a sample of 8 spiral galaxies from the THINGS survey and found similar flaring \hi\ disks. The \hi\ scale height increases from a few hundred parsecs at the center to $\sim1-2$\,kpc at the edge. The median axial ratio was found to be 0.1, indicating much thinner \hi\ disks than that in dwarf galaxies. In addition, very low axial ratios were found for three galaxies, suggesting that they could be potential super-thin galaxies.

\subsection{Turbulence in the ISM}
\label{sec:emission_turbulence}

Turbulence in the ISM is known to be a driving factor behind morphology, dynamics and star formation of galaxies \citep{elmegreen2004}. Observations in the MW \citep{crovisier1983,green1993} and dwarf and spiral galaxies \citep{begum2006,dutta2008,dutta2009,dutta2013} show that the column density power spectrum assumes power law over large length scale ranges. \citet{nandakumar2020} implemented a variant of column density and line-of-sight velocity power spectrum estimator introduced in \citet{duttap2016}, that uses the visibilities measured by interferometers directly. Combining VLA and uGMRT observations, they report the existence of a single power law power spectrum in NGC~5236 over two decades of length scales. They also report the first ever measurement of the line-of-sight velocity power spectrum in any external spiral galaxy. Using results from simulations and the measured slope, they conclude that the large-scale turbulence is generated by compressive forcing that probably originated from self gravity and instabilities in the disk. Since the energy cascade from the large-scale is of the same order as that by instabilities in the ISM through supernovae feedback, this has significant implication on the evolution of the disk morphology and the star formation efficiency.

\subsection{Scaling relations}
\label{sec:emission_scaling}

We discuss here some of the scaling relations between global properties of galaxies involving \hi\ gas that provide us with insights into the complex processes governing galaxy formation and evolution. 

\subsubsection{\hi\ size--mass relation:} 
\label{sec:emission_sizemass}

The scaling relation between \hi\ mass (M$_{\rm HI}$) and diameter of the \hi\ disk (D$_{\rm HI}$) defined at a \hi\ surface density of 1 M$_{\odot}$ pc$^{-2}$ was first parameterised by \citet{broeils97} as:

\begin{equation}
    \mathrm{ log(D_{\rm HI}) \ = \ 0.5 \ log(M_{\rm HI}) - 3.32 }
    \label{eqn_sizemass}
\end{equation}

\noindent
where D$_{\rm HI}$ is in kpc and M$_{\rm HI}$ is in M$_{\odot}$ pc$^{-2}$. Subsequent \hi\ surveys have confirmed that the relation holds true for a wide range of galaxy masses and morphologies such as large spirals \citep[e.g][]{verheijen01, swaters02, wang13, lelli16, ponomareva16}, late-type dwarf galaxies \citep[e.g][]{swaters02, begum2008a, lelli16}, early-type spirals \citep[e.g.][]{noordermeer05} and irregulars \citep[e.g.][]{lelli16}. The ultra-diffuse galaxies that have extreme ratios of stellar mass to stellar scale length and were discovered in \hi\ surveys of isolated environments \citep[e.g.][]{leisman17}, were also found to follow the \hi\ size--mass relation \citep{leisman17, gault21}. We expect galaxies residing in groups and clusters to lose gas due to various environmental processes such as ram pressure stripping. However, galaxies in groups and clusters were also shown to follow the above relation as long as their discs are not too disrupted and the diameter can be traced out to 1 M$_{\odot}$ pc$^{-2}$ \citep{verheijen01, chung2009}.
\citet{wang16} have obtained a remarkably tight \hi\ size--mass relation with a scatter of $\sim$0.06 dex, based on \hi\ sizes for more than 500 galaxies from various projects, ranging over five decades in \hi\ mass. They find that the scatter does not change as a function of galaxy luminosity, \hi\ richness or morphology, which indicates a constant average \hi\ surface density within D$_{\rm HI}$ for most galaxies. \citet{stevens19} analytically derived the \hi\ size--mass relation of galaxies and showed that the relation can be used to constrain galaxy formation and evolution theories, where the success of any model or simulation is based on its ability to accurately reproduce the scatter, slope and zero point of this relation.

\subsubsection{Baryonic Tully-Fisher relation:} 
\label{sec:emission_tullyfisher}

The Tully-Fisher (TF) relation is a tight correlation between the luminosity of a disk galaxy and its rotational velocity \citep{tully77}. This relation has been found to hold for rotating galaxies of various morphologies \citep[e.g.][]{chung02, courteau03, denheijer15, karachentsev17}, in different environments \citep[e.g.][]{willick99, abril21}, and over a large wavelength range \citep[e.g.][]{verheijen01, ponomareva17}. Hence, the relation has became a major tool to place tight constraints on galaxy formation and evolution models \citep[e.g.][]{navarro00, maccio16}.

However, this relation breaks down for velocities smaller than 100\,\kms\ \citep{mcgaugh00}. This regime is mainly dominated by low-mass galaxies, where the cold gas mass is comparable to or greater than the stellar mass. When the stellar mass is replaced with the baryonic mass, one recovers a single, tight linear relation that spans over $\sim$5\,dex in baryonic mass \citep{begum2008b, mcgaugh12, lelli16b, lelli19}. This is called as the baryonic TF relation (BTFr), the most fundamental form of the TF. The definition of circular velocity is shown to significantly alter the resulting BTFr \citep{brook16, ponomareva17, lelli19}. \citet{lelli19} analyzed a sample of 175 spirals based on different velocity definitions such as W$_{\rm 50}$ -- rotation velocity derived from the integrated \hi\ spectrum, V$_{\rm flat}$ -- velocity measured at the flat part of the rotation curve, and V$_{\rm max}$ -- maximum velocity measured from the rotation curve. They found that the tightest BTFr and steepest slope is given by the V$_{\rm flat}$ method. The intrinsic scatter of the BTFr is lower than the predictions from $\Lambda$CDM cosmology \citep{dutton12} and the BTFr slope with V$_{\rm flat}$ is higher than the slope expected in $\Lambda$CDM models \citep{mcgaugh12}. The galaxy formation and evolution models predict that the BTFr residuals correlate with galaxy size and surface brightness \citep{desmond15}. In contrast to this, the observed BTFr residuals do not show any correlation with galaxy radius \citep{lelli19}. A curvature at the low-mass end of BTFr was predicted by some semi-analytical galaxy formation models \citep{trujillo11, desmond12}, but \citet{iorio17} find that dwarf irregulars lie exactly on the BTFr with no evidence for curvature. However, \hi-rich low surface brightness galaxies classified as ultra-diffuse galaxies were found to be clear outliers from the BTFr, with circular velocities much lower than galaxies with similar baryonic mass \citep{pina19}. Further, \citet{ponomareva21} used a sample of 67 galaxies to study the BTFr over a period of $\sim$1 billion year (0 $\leq$ z $\leq$ 0.081) and found that all the galaxies are consistent with the same relation independent of redshift.

\subsection{Stacking studies}
\label{sec:emission_stacking}

Two observational approaches have been used to quantify the amount of \hi\ in the Universe, viz. using \hi\ 21-cm emission at $z\lesssim$0.2 and using \lya\ absorption at $z\ge$2 (see Section~\ref{sec:absorption_intervening_lya}). At intermediate redshifts, there are limited observational probes of the \hi\ content leading to the cosmic mass density of \hi\ being poorly constrained. Stacking the \hi\ 21-cm emission from a large sample of galaxies with accurate redshifts can be used to measure the average \hi\ content of different galaxy populations at intermediate redshifts. However, despite efforts there were no statistically significant detections of stacked \hi\ 21-cm emission at $z\gtrsim$0.2 \citep[e.g.][]{lah2007,kanekar2016,rhee2018}. Recently, \citet{bera2019} detected a stacked \hi\ 21-cm emission signal at $\approx$7$\sigma$ significance from a sample of 445 blue star-forming galaxies at $0.2<z<0.4$ using uGMRT. Their results implied an average \hi\ mass of $5\times10^9$\,\msun\ and no significant evolution in the cosmic \hi\ mass density over $z\approx$0-0.4. \citet{chowdhury2020b} reported the first detection (at $\approx$4.5$\sigma$ significance) of stacked \hi\ 21-cm emission at $z\approx$1 from 7653 star-forming galaxies over $0.74\le z\le1.45$ using uGMRT. They estimated an average \hi\ mass of $10^{10}$\,\msun\ that is similar to the average stellar mass of the sample. Combining with the average SFR estimated from stacked 1.4\,GHz continuum emission, they derived an average \hi\ depletion time of $\approx$1.5\,Gyr. The cosmic \hi\ mass density estimated from this study is consistent within the uncertainties with the measurements at $z\le$1, but is lower by a factor of $\approx$2 than the measurements at $z\ge$2. The above study was followed by an independent detection of stacked \hi\ 21-cm emission at $z\approx$1.3 that led to similar results \citep{chowdhury2021}.

\section{\hi\ 21-cm absorption}
\label{sec:absorption}

\hi\ 21-cm absorption studies provide significant and complementary insights into the atomic gas in galaxies. \hi\ absorption has both advantages and disadvantages compared to \hi\ emission. The main disadvantage of \hi\ emission observations is that they are limited by the distance of the emitting source due to the weakness of the 21-cm transition. \hi\ emission can be observed only in the relatively local Universe ($z<0.1$) in reasonable time with the capabilities of the current radio telescopes. On the other hand, the detectability of \hi\ absorption depends only on the strength of the background radio continuum source and the \hi\ gas cross-section of the galaxy. Besides being a redshift-independent probe of the atomic gas in galaxies, \hi\ absorption can be observed at very high spatial resolution, i.e., $\sim$milliarcsec (mas) scales, with Very Long Baseline Interferometry (VLBI) if there is a bright enough background source, which is not possible with \hi\ emission. 

A disadvantage intrinsic to \hi\ absorption technique is that it provides information about the atomic gas only in the regions where there is sufficiently bright background radio continuum emission, which may not extend over the full gaseous structure being probed. Another disadvantage of \hi\ absorption studies is that estimates of the neutral hydrogen column density (\nhi) is dependent on the excitation or spin temperature (\ts) of the 21-cm transition and the fraction of the background radio continuum source that is covered by the absorbing gas (\fc) in case of unresolved source, and these parameters are usually difficult to constrain observationally.
The optical depth integrated over the \hi\ 21-cm absorption line profile in velocity space (\taudv; \kms) is directly related to \nhi\ (\cms) and inversely related to \ts\ (K) of the gas, as follows \citep{field1959,kulkarni1988}:

\begin{equation}
 N(\hi) ~ [{\rm cm}^{-2}] = 1.823 \times 10^{18} ~ T_{\rm s} ~ [K] \int{\tau {\rm dv}}  ~ [{\rm km~s}^{-1}],
\label{eqn_absorption1}
\end{equation}

\noindent
where \ts\ represents the effective spin temperature, which is the \nhi-weighted harmonic mean spin temperature of different gas clouds along the line of sight that give rise to the absorption. The 21-cm optical depth corrected for covering factor (\fc) is given by, 

\begin{equation}
 \tau = -{\rm log}[1 - \frac{\Delta S}{(f_{\rm c} S)}],
\label{eqn_absorption2}
\end{equation}

\noindent
where $S$ is the flux density of the background radio continuum source and $\Delta S$ is the depth of the absorption line. For optically thin gas, Equation~\ref{eqn_absorption1} reduces to,

\begin{equation}
 N(\hi) ~ [{\rm cm}^{-2}] = 1.823 \times 10^{18} ~ \frac{T_{\rm s} ~ [K]}{f_{\rm c}} \int{\frac{\Delta S}{S} {\rm dv}} ~ [{\rm km~s}^{-1}].
\label{eqn_absorption3}
\end{equation}

It is possible to put limits on \ts\ by comparing \nhi\ obtained from \hi\ 21-cm absorption with that obtained independently from \hi\ 21-cm emission \citep[e.g.][]{kanekar2001a,dutta2016}, \hi\ \lymana\ absorption \citep[e.g.][]{srianand2012,kanekar2014a} or X-ray absorption \citep[e.g.][]{allison2015,moss2017} of the same source. Note that each of the above methods have their limitations. For example, \hi\ 21-cm emission and absorption can be detected simultaneously only at $z<0.1$ and even then the angular scales probed by emission observations will be much larger than that probed by absorption. Comparison of \hi\ 21-cm absorption with \lya\ absorption in the ultraviolet (UV) or optical spectra of the background source is affected by uncertainties arising from the assumptions that the radio and UV/optical sightlines are aligned and they trace the same gas phase. Similarly, comparison with the total column density of hydrogen from X-ray absorption to obtain an upper limit on \ts\ is based upon the assumption that the radio and X-ray spectra probe similar sightlines through the gas.

The covering factor, \fc, can be directly constrained using high spatial resolution (mas- or pc-scale) VLBI redshifted \hi\ 21-cm line observations. However, the limited frequency coverage of current VLBI facilities restricts such observations to few sources at low redshifts \citep[e.g.][]{morganti2013,srianand2013,biggs2016}. Alternatively, we can indirectly constrain \fc\ from the core fraction by comparing the flux density of the core component of the background radio source at high spatial resolution from VLBI images with the total flux density at the lower spatial resolution (arcsec- or kpc-scale) at which the absorption is observed \citep[e.g.][]{kanekar2009a,gupta2012,srianand2012}. This method assumes that the core of the radio source is spatially aligned with the optical source.

\begin{figure*}[!t]
\includegraphics[width=1.0\textwidth]{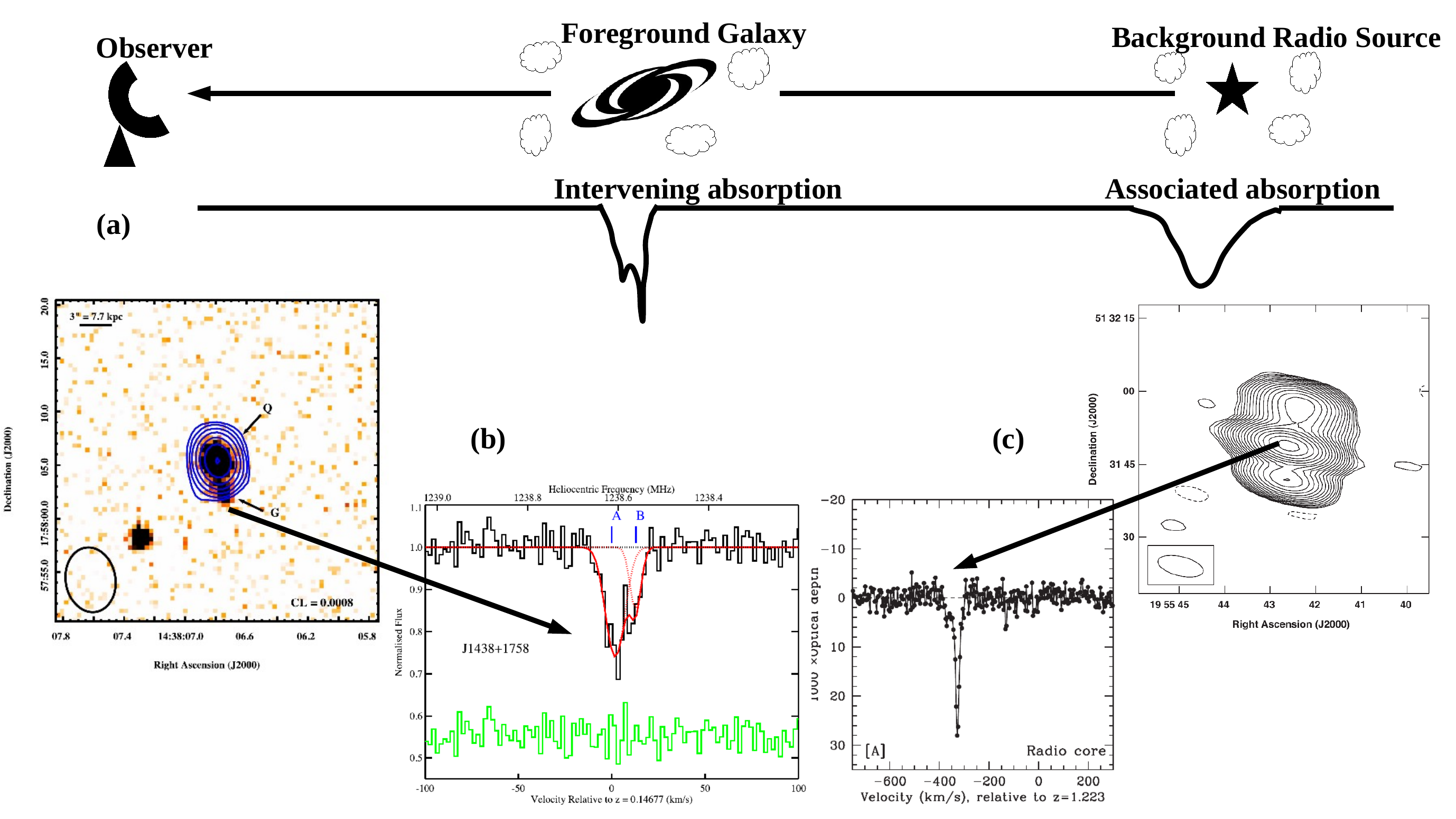}
\caption{
(a) Cartoon showing how \hi\ 21-cm absorption can be used to probe gas in radio-bright sources (associated absorption) and gas around foreground galaxies (intervening absorption).
(b) Example of intervening \hi\ 21-cm absorption detected using GMRT from a $z=0.15$ galaxy (marked by ``G'') at a projected separation of $\approx8$\,kpc from a background radio quasar (marked by ``Q'') \citep[reproduced from][]{dutta2017a}. The contours show the 1.4\,GHz radio continuum emission.
(c) Example of associated \hi\ 21-cm absorption detected using GMRT against the core of an extended radio source at $z=1.2$ \citep[reproduced from][]{aditya2017}. The contours show the 640\,MHz continuum emission from the radio source. 
}
\label{fig_absorption}
\end{figure*}

\hi\ 21-cm absorption line technique can be used to study both the gas located in/around foreground galaxies that lie along the line of sight to a background radio continuum source \citep[i.e., intervening absorption, see e.g.][]{dutta2019a}, and the gas located in the radio continuum source itself, typically an Active Galactic Nucleus or AGN \citep[i.e., associated absorption, see e.g.][]{morganti2018}. We discuss how intervening \hi\ 21-cm absorption has been used to study the physical properties of the neutral gas phase in different types of galaxies in Section~\ref{sec:absorption_intervening} and how associated \hi\ 21-cm absorption has been used to probe AGN feeding and feedback and AGN-galaxy co-evolution in Section~\ref{sec:absorption_associated}. Examples of intervening and associated \hi\ 21-cm absorption lines are shown in Fig.~\ref{fig_absorption}.

We note that in addition to studying the \hi\ gas in galaxies and AGNs, \hi\ 21-cm absorption lines in conjunction with OH 18-cm absorption lines and metal absorption lines (assuming they trace the same gas) detected in high-resolution optical spectra can be used to place stringent constraints on the variation in the fundamental constants of physics such as combination of the fine structure constant, the proton-to-electron mass ratio and the proton gyromagnetic ratio over cosmological scales \citep[e.g.][]{wolfe1976,chengalur2003,srianand2010,kanekar2012,rahmani2012,gupta2018b}. Further, precise redshifts obtained from \hi\ 21-cm absorption line observations in multiple epochs can be used to constrain the secular redshift drift or the cosmic acceleration \citep{darling2012,jiao2020}. However, a detailed discussion of such studies is beyond the scope of this article.

\subsection{Intervening \hi\ absorption}
\label{sec:absorption_intervening}

\subsubsection{\hi\ absorption in the Milky Way:}
\label{sec:absorption_intervening_mw}

Observations of \hi\ 21-cm absorption have been instrumental in shedding light on the physical conditions such as the thermal state of the neutral gas. In the cold neutral medium (CNM), the spin temperature is coupled with the kinematic temperature \citep[\tk;][]{field1959,bahcall1969,roy2006}, whereas in the warm neutral medium (WNM), \ts\ can provide an upper limit on \tk\ \citep{field1958,liszt2001,kim2014}. Additionally, the linewidths of individual \hi\ 21-cm absorption components detected in high-velocity resolution spectra can place constraints on the gas \tk\ \citep{lane2000a,kanekar2001b,koley2019}. There have been extensive studies, based on \hi\ 21-cm absorption, of the temperature, column density and filling fraction of the neutral gas phase in the MW \citep[e.g.][]{heiles2003,kanekar2003b,mohan2004,begum2010,roy2013,murray2015,murray2018,patra2018}. Combining with \hi\ emission observations, a key result from the above studies is that a significant fraction ($\approx20-50$\%) of the \hi\ gas in the MW is present in the thermally unstable regime with \tk\ $\approx500-5000$\,K, i.e. outside the stable range predicted by steady-state two-phase models of the neutral gas. This suggests that dynamical processes such as turbulence, supernovae and magnetic fields could be driving the gas from the stable to the unstable phase \citep{audit2005,deavillez2005}. 

To probe the effect of the above processes on gas dynamics in the ISM, there have been \hi\ 21-cm absorption studies of the magnetic field and turbulence in the MW \citep[e.g.][]{heiles2004,heiles2005,roy2008}. The Zeeman effect in radio-frequency spectral lines is one of the most direct methods to measure the strength of the magnetic field \citep{troland1982}. While Zeeman-splitting of the \hi\ 21-cm absorption line has been detected in only one extraglactic source \citep{kazes1991,sarma2005}, using such measurements in the MW, \citet{heiles2005} found that magnetic field dominates thermal motions in the CNM, and that magnetism and turbulence are in approximate equipartition. 

\subsubsection{\hi\ absorption in galaxies at small projected separations from quasars:}
\label{sec:absorption_intervening_qgp}

As motivated earlier, it is challenging to detect the diffuse gas around galaxies in emission as we go beyond the local Universe, but it can be detected in absorption towards a bright background source, e.g. AGN or quasar. Depending on the location of the background source with respect to the foreground galaxy, we can probe different components of the galaxy such as the ISM in the galactic disc, extended \hi\ disc, outflows, inflows, high velocity clouds and tidal streams in the halo or CGM around the galaxy. Although absorption line technique probes gas only along the line of sight, we can map the gas distribution in a statistical manner provided there is a sizeable sample of pairs of background sources and foreground galaxies. Applying this method, there have been numerous studies of the different gas phases around galaxies. Such studies have shown that the gaseous haloes or the CGM is a crucial component of galaxies and observations of different gas phases in the CGM can place useful constraints on models of galaxy formation and evolution \citep[see for a review][]{tumlinson2017}. Due to the inverse dependence of its optical depth on \ts, \hi\ 21-cm absorption is an excellent tracer of the cold neutral gas in the ISM and CGM of galaxies. 

There have been several surveys of \hi\ 21-cm absorption from foreground galaxies at $z<0.4$ that were selected to lie at small project separations or impact parameters ($\lesssim50$\,kpc) from a background radio-loud quasar \citep[e.g.][]{carilli1992,gupta2010,borthakur2011,borthakur2016,zwaan2015,reeves2016,dutta2017a}. \citet{dutta2017a} found a weak ($\approx3\sigma$) anti-correlation between \hi\ 21-cm optical depth and impact parameter from galaxies. The covering fraction or incidence of \hi\ 21-cm absorption similarly decline with increasing impact parameter, with an average incidence of $\approx20$\% within 30\,kpc for an optical depth sensitivity of \taudv\ $=0.3$\,\kms. The optical depth and incidence also tend to be larger near the galaxy major axis, suggesting that the majority of the absorbing gas is co-planar with the extended \hi\ discs of galaxies. This is further supported by the properties of metal absorption lines of \caii\ and \nai\ detected in the optical spectra of the background quasars, which suggest that most of the \hi\ absorbers detected around low-$z$ galaxies trace the diffuse extended \hi\ discs rather than the dusty star-forming discs of galaxies. On the other hand, the \hi\ 21-cm optical depth and incidence are not found to depend significantly on the galaxy properties such as luminosity, stellar mass, colour and SFR. 

\subsubsection{\hi\ absorption in \mgii-selected galaxies:}
\label{sec:absorption_intervening_mgii}

It is difficult to extend the above type of studies to $z\gtrsim0.5$, where there is a dearth of large, wide-field galaxy spectroscopic surveys. However, absorption lines can be used to select galaxies in a luminosity-unbiased way. In particular, the \mgii\ doublet lines, $\lambda\lambda$2796, 2803, have been used extensively to probe the gaseous haloes of $z\lesssim2$ galaxies \citep[e.g.][]{bergeron1986,steidel1994,churchill2000,chen2010,nielsen2013,rubin2018,dutta2020b,dutta2021}. Moreover, strong \mgii\ absorption lines, i.e. those with rest-frame equivalent width of \mgii\ $\lambda$2796 line \wmgii\ $\ge1$\,\AA, have been found to trace gas with high \hi\ column densities \citep[\nhi\ $\gtrsim10^{19-20}$\,\cms][]{rao2006}. Therefore, \hi\ 21-cm observations of \mgii\ systems can be used to trace the neutral gas at $z<1.5$, a crucial redshift range over which the global star formation rate density declines and where the \lymana\ line cannot be observed from the ground.

Taking advantage of large samples of \mgii\ absorbers detected in the optical spectra of quasars from surveys such as the Sloan Digital Sky Survey \citep[SDSS;][]{york2000}, there have been several searches of \hi\ 21-cm absorption in \mgii\ systems \citep[e.g.][]{briggs1983,lane2000b,kanekar2009b,gupta2009,gupta2012,dutta2017b,dutta2017c}. Based on these studies, the average detection rate of \hi\ 21-cm absorption in samples of strong \mgii\ systems is $\approx10-20$\% for a sensitivity of \taudv\ $=0.3$\,\kms. This incidence is found to remain constant within the uncertainties over the redshift range $z\approx0.3-1.5$ \citep{gupta2012,dutta2017c}. Similarly, the redshift evolution in the number density per unit redshift of \hi\ 21-cm absorbers in strong \mgii\ systems is found to be consistent with that of a non-evolving population, although these results currently are limited by large uncertainties due to small number statistics. Next, the detection rate of \hi\ 21-cm absorption is found to be higher towards background quasars with flat or inverted spectral index and small linear sizes, indicating that the absorbing gas is patchy with characteristic size $\lesssim100$\,pc \citep{gupta2012}. Further, the \hi\ 21-cm absorption detection rate is shown to increase with the equivalent width of \mgii\ and \feii\ \citep{gupta2012,dutta2017b}.  Equivalent width ratio cuts of \mgii, \mgi\ and \feii\ that increase the probability of detecting high \nhi\ gas also lead to higher detection rate of \hi\ 21-cm absorption. In addition, stronger \hi\ 21-cm absorption and higher detection rate is found towards quasars which are more reddened due to dust in the absorbing gas \citep{dutta2017b}. All these results point towards a close link between metals, dust and cold \hi\ gas in absorption-selected galaxies at $z<1.5$.

\subsubsection{\hi\ absorption in \lya-selected galaxies:}
\label{sec:absorption_intervening_lya}

Apart from \mgii, strong \hi\ \lya\ absorption in the spectra of background quasars is a well-established probe of the gas around galaxies. Specifically, damped \lymana\ absorbers (DLAs) with \nhi\ $\ge2\times10^{20}$\,\cms\ and sub-DLAs with \nhi\ $\approx10^{19}-2\times10^{20}$\,\cms, are found to be the major reservoirs of \hi\ gas over $2\le z \le4$ \citep{wolfe2005,peroux2005,prochaska2005,noterdaeme2012}. While the number ($<100$) of (sub-)DLAs detected at $z<1.5$ is relatively less due to the requirement of UV spectra of quasars \citep[e.g.][]{rao2017}, thousands of (sub-)DLAs have been identified at $z>1.8$ from SDSS \citep[e.g.][]{noterdaeme2012}. There have been several efforts to connect the properties of (sub-)DLAs (column density, metallicity) with the properties (stellar mass, atomic and molecular gas mass, SFR) of the associated galaxies \citep[e.g.][]{rao2011,peroux2012,fumagalli2015,chengalur2015,srianand2016,krogager2017,kanekar2018a,kanekar2018b,mackenzie2019}. From such studies, the success rate of detecting DLA galaxies is found to be higher for the high metallicity systems, and DLAs are generally thought to be associated with faint, gas-rich galaxies. 

Complementary to the above, \hi\ 21-cm absorption studies of DLAs detected towards radio-loud quasars offer important insights into the physical conditions of the neutral gas in galaxies \citep[e.g.][]{wolfe1985,carilli1996,chengalur2000,kanekar2003a,curran2010,srianand2012,kanekar2014a}. The above studies indicate that fraction of cold gas traced by \hi\ 21-cm absorption in $z\ge2$ DLAs is small, with a detection rate of $\approx20$\% for a sensitivity of \taudv\ $=0.3$\,\kms. Estimates of spin temperature of the gas from comparison of \nhi\ from \lya\ absorption and optical depth from 21-cm absorption imply that a significant fraction of the \hi\ gas in $z\ge2$ DLAs traces the WNM \citep{srianand2012,kanekar2014a}. The relatively low CNM fraction ($\approx10-20$\%) in $z\ge2$ DLAs inferred from \hi\ 21-cm absorption studies is consistent with similar low molecular (H$_2$) fraction seen in such DLAs \citep{noterdaeme2008}. Further, the anti-correlation observed between the gas metallicity and \ts\ in DLAs \citep{kanekar2009c} also suggests that the high \ts\ values obtained in $z\ge2$ DLAs arise due to larger WNM fractions in the low-metallicity environment typical in such DLAs, likely due to fewer radiative cooling routes. The spin temperature shows a redshift evolution, with a larger number of $z>2.4$ DLAs having \ts\ $\gtrsim1000$\,K leading to the \ts\ distribution of $z>2.4$ DLAs being significantly different than that of $z<2.4$ DLAs and also that of the MW \citep{kanekar2014a}. In addition, the incidence of \hi\ 21-cm absorption in DLAs is found to decline by a factor of $\approx3$ from $z<1$ to $z>2$. The above results are indicative of a decline in the cold gas fraction in galaxies with cosmic time.

\subsubsection{Spatially-resolved studies of \hi\ gas in galaxies:}
\label{sec:absorption_intervening_highres}

\begin{figure*}[!t]
\centering
\includegraphics[width=1.0\textwidth,trim={0 0.5cm 0 0},clip]{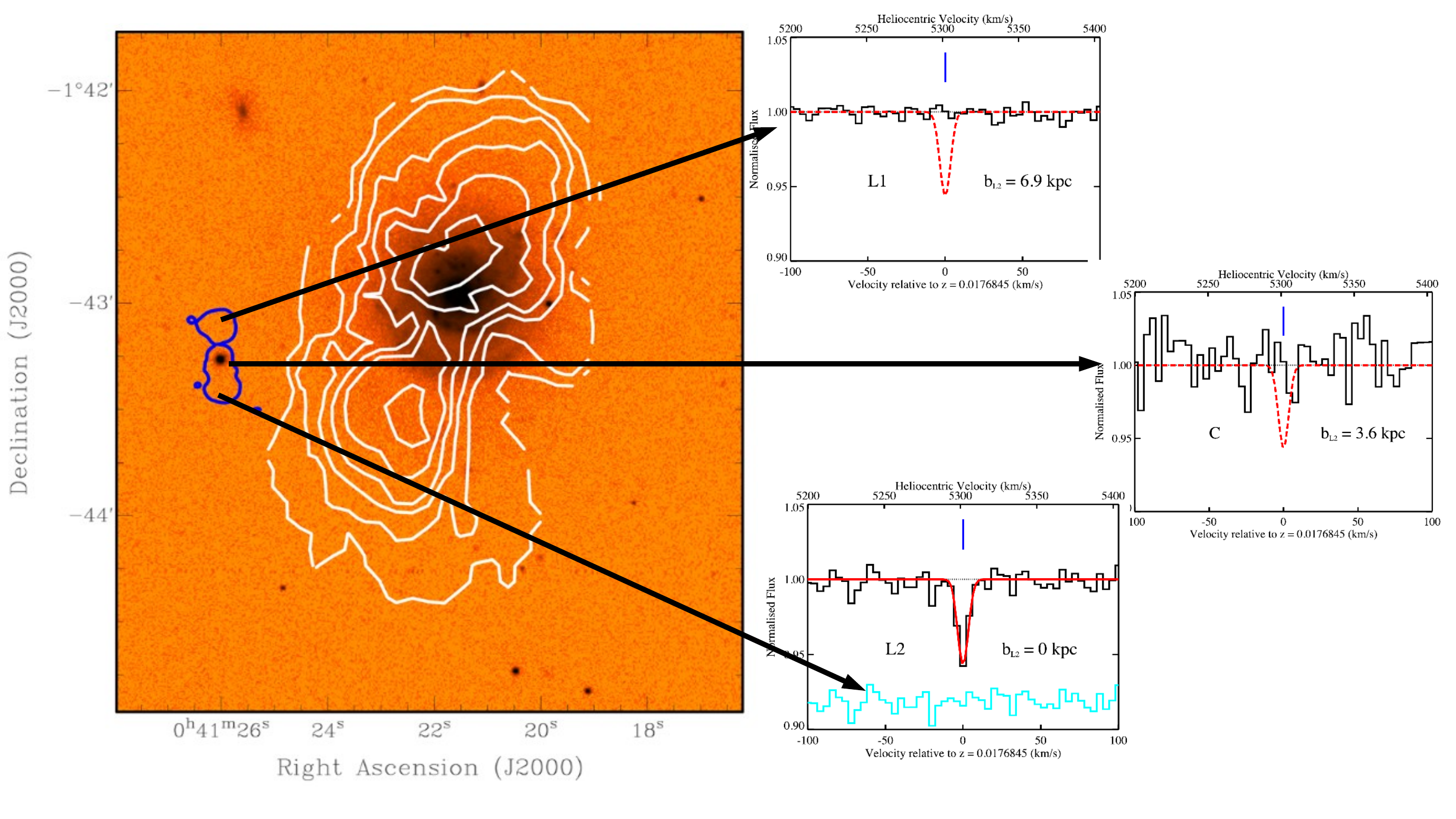}
\caption{GMRT \hi\ 21-cm absorption spectra towards the lobes and core of a background extended radio source from the study of \citet{dutta2016}. \hi\ 21-cm absorption is detected towards the southern lobe from the extended \hi\ disc of a $z\approx0.02$ galaxy, but is not detected towards the core or the northern lobe. The white contours represent the \hi\ emission detected from the galaxy and the blue line represents the outermost contour of the 1.4\,GHz continuum emission of the radio source that extends over 7\,kpc at the redshift of the galaxy.}
\label{fig_hiabs_resolved}
\end{figure*}

The scarcity of strong extended background radio sources means that there are only few spatially-resolved studies of \hi\ gas in absorption. \citet{dutta2016} reported detection of \hi\ 21-cm absorption from the extended \hi\ disc of a $z\approx0.02$ galaxy over an area of $\approx2$\,kpc$^2$ against one of the lobes of a background radio source. The optical depth is observed to vary by more than a factor of 7 over the extent of the radio source of $\approx7$\,kpc at the redshift of the galaxy (see Fig.~\ref{fig_hiabs_resolved}). VLBI spectroscopic studies have been possible in case of few $z\le0.1$ \hi\ 21-cm absorbers \citep{lane2000a,keeney2005,borthakur2010,srianand2013,biggs2016,gupta2018a}. These studies infer sizes of the coherent \hi\ gas structures to be $\gtrsim20-30$\,pc. Variations in the \hi\ optical depth by a factor of $\approx2-5$ over scales of $\lesssim10-100$\,pc have been observed in denser gas in galaxies probed by extended background radio emission \citep{srianand2013,biggs2016}. Variations in the \hi\ optical depth with time in few cases also suggest that the \hi\ gas is patchy over pc-scales \citep{wolfe1982,kanekar2001c,allison2017}. The small-scale patchiness of the \hi\ 21-cm absorbing gas is further supported by VLBI imaging observations of the background radio sources and observations of H$_2$ absorption along the same sightlines \citep[e.g.][]{srianand2012,gupta2012,dutta2015}.

\subsection{Associated \hi\ absorption}
\label{sec:absorption_associated}

The majority of bright extragalactic radio sources are powered by radio jets produced in the central AGN of galaxies \citep{condon1998}. While the physical mechanisms that trigger the AGN activity is an area of ongoing research, it is believed that gas funnelled to the central regions of a galaxy either via secular processes or interactions and mergers play a key role in activating the AGN activity \citep{hopkins2008}. Outflows induced by radio jets from the AGN can either ignite or suppress star formation activity in galaxies, i.e. both positive and negative AGN feedback can strongly impact the evolution of the host galaxies and the environment \citep{schawinski2006,maiolino2017}. It is thus evident that gas flows play a vital role in the co-evolution of AGN and galaxies \citep{fabian2012}. \hi\ 21-cm absorption is a well-utilised tool to study the distribution and kinematics of the neutral gas in centres of active galaxies. It can be used to study the cold ISM gas in the vicinity of AGN, and probe fuelling and feedback processes associated with AGN in the form of neutral gas inflows and outflows \citep[e.g.][]{hota2005,teng2013,morganti2018}.

\subsubsection{\hi\ absorption in low-$z$ AGN:}
\label{sec:absorption_associated_lowz}

There have been numerous studies of \hi\ 21-cm absorption in different samples of radio-loud AGNs at low redshifts. One of the first systematic surveys of associated \hi\ 21-cm absorption was conducted by \citet{vangorkom1989}, who found evidence for higher occurrence of absorption lines redshifted relative to the systemic velocity. In contrast to this, subsequent studies have found higher occurrence of blueshifted absorption \citep[e.g.][]{vermeulen2003}, indicating that disentangling gas flows in the circumnuclear regions is complex. The overall detection rate of \hi\ 21-cm absorption associated with $z<0.4$ radio AGNs is found to be $\approx20-30$\% \citep[e.g.][]{morganti2001,vermeulen2003,curran2006,gupta2006,allison2012,chandola2013,gereb2015,maccagni2017,glowacki2017,murthy2021}. 

However, the detection rate of \hi\ 21-cm absorption is found to vary across different classes of AGNs. The properties of the absorbing \hi\ gas are found to depend on the radio continuum emission, with higher incidence ($\approx30-60$\%) in compact young radio sources (compact steep-spectrum and giga-Hertz peaked spectrum sources) compared to extended radio sources ($\lesssim15$\%), due to a high gas covering factor \citep[e.g.][]{pihlstrom2003,vermeulen2003,gupta2006,chandola2013,maccagni2017}. Compact radio sources also show higher frequency of blueshifted and broad or asymmetric absorption lines implying the presence of unsettled gas \citep[e.g.][]{gereb2015,glowacki2017}. The physical conditions in the ISM of the host galaxy such as presence of dust are additionally found to influence the occurrence of associated \hi\ absorption. Based on their mid-infrared colours, dusty gas-rich radio AGN are found to exhibit higher incidence ($\approx40-70$\%) of \hi\ absorption compared to dust-poor sources \citep[e.g.][]{chandola2017,maccagni2017}. While low-excitation radio galaxies (LERGs) are found to yield overall lower detection rate of \hi\ absorption compared to high-excitation radio galaxies (HERGs) likely due to suppressed star formation, LERGs with compact radio emission and bright mid-infrared colours give rise to high detection rate \citep[$\approx70$\%;][]{chandola2017}.

Further, radio galaxies are found to have two times higher \hi\ absorption detection rate compared to radio quasars \citep{gupta2006}, consistent with the unification model where the circumnuclear structure obscures a radio galaxy or type-2 AGN, while quasars or type-1 AGN are unobscured. However, high incidence ($\approx80$\%) of \hi\ absorption has been observed in radio-selected red quasars that likely represent early dust-obscured phase of AGN \citep{carilli1998a}. In addition, \hi\ absorption is found to be highly prevalent ($\approx84$\%) in radio AGN present in galaxies undergoing mergers \citep{dutta2018,dutta2019b}. Radio-loud mergers show significantly stronger ($\approx5\times$) \hi\ absorption and also higher ($\approx3\times$) occurrence of redshifted absorption compared to non-merging radio AGNs. This indicates that the merging process is efficient in funnelling large quantities of neutral gas to the galaxy centres and that it could be linked with triggering of the AGN activity.

\subsubsection{\hi\ absorption in high-$z$ AGN and redshift evolution:}
\label{sec:absorption_associated_highz}

\begin{figure*}[!t]
\centering
\includegraphics[width=0.45\textwidth]{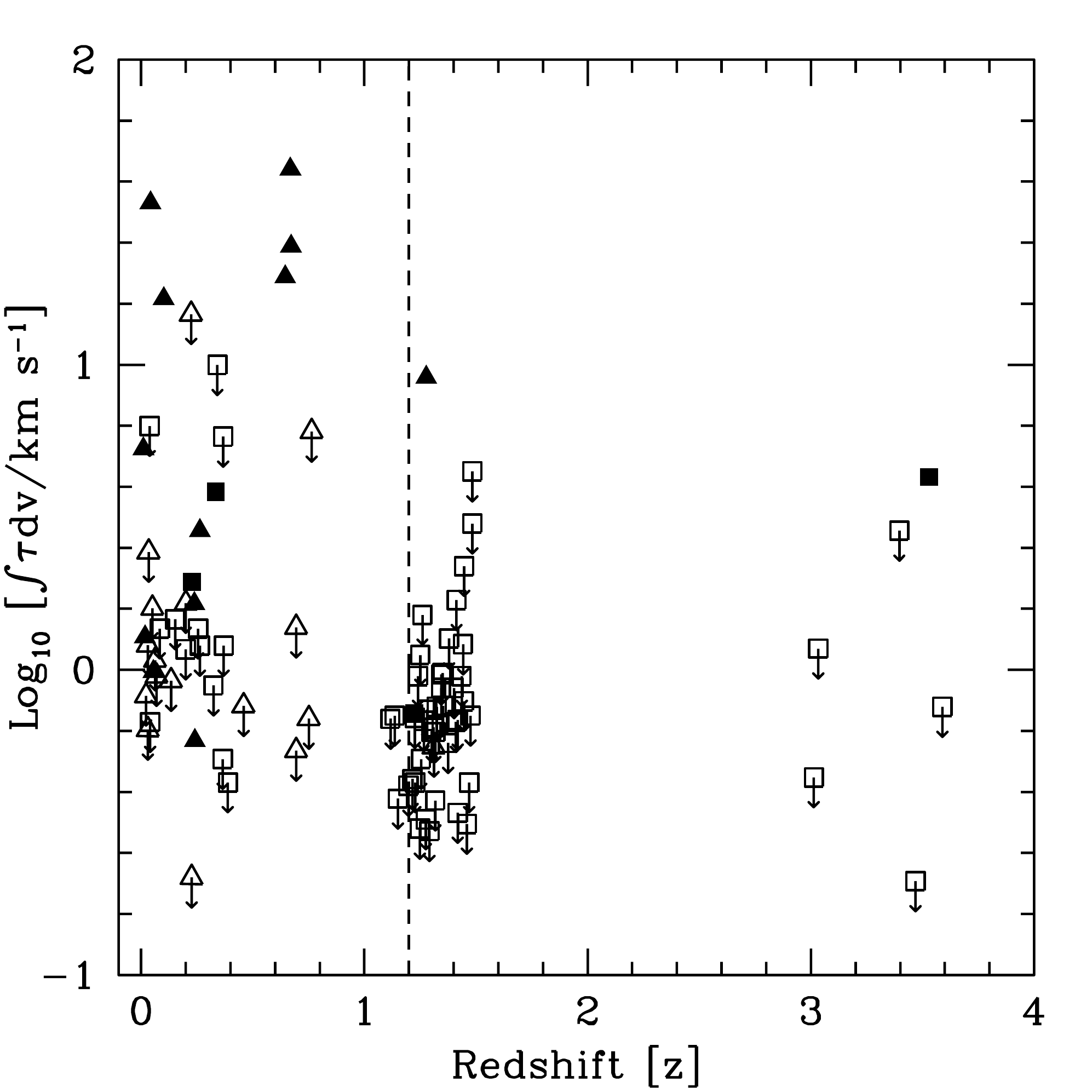}
\includegraphics[width=0.45\textwidth]{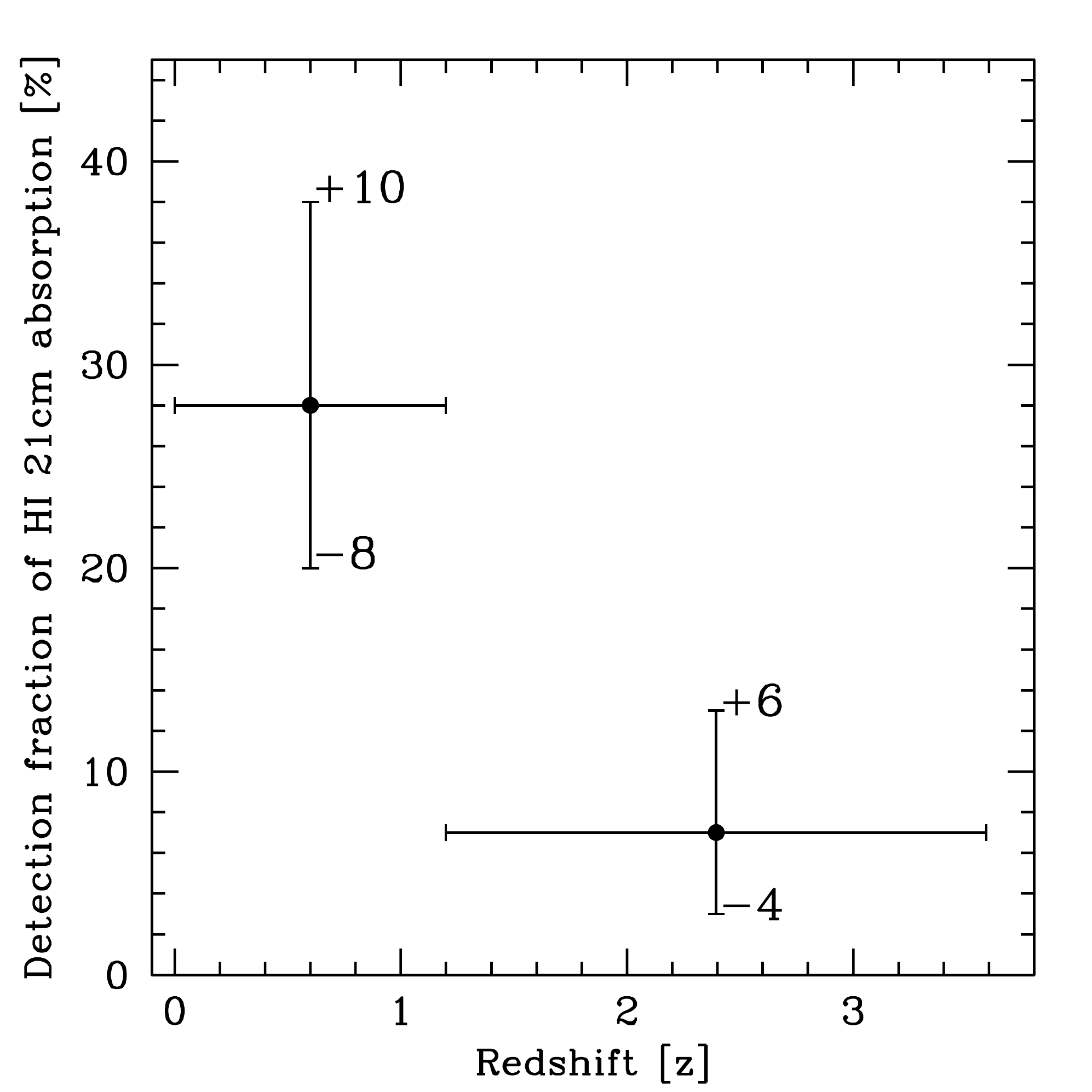}
\caption{The velocity-integrated \hi\ 21-cm optical depth (left) and detection rate of \hi\ 21-cm absorption (right) as a function of redshift in flat-spectrum radio sources from the study of \citet{aditya2018a}.}
\label{fig_hiabs_zevolution}
\end{figure*}

While the above studies have found more than 150 associated \hi\ 21-cm absorbers at $z<1$, endeavours to detect these absorbers at $z>1$ have met with low detection rates \citep[$\lesssim10$\%; e.g.][]{curran2008,curran2013a,aditya2016,aditya2018a,aditya2018b,grasha2019}. There are currently only 12 detections of associated \hi\ 21-cm absorption at $z>1$ \citep{uson1991,moore1999,ishwarachandra2003,curran2013b,aditya2017,aditya2018a,aditya2021,dutta2020a,chowdhury2020a,gupta2021b}. \citet{curran2013a} argued that a high UV luminosity of the AGNs at high redshifts is likely the dominant cause for the low detection rate at high redshifts. In a flux-limited sample, the AGNs at high redshifts would have systematically higher UV luminosities owing to the Malmquist bias. The strong UV radiation would ionise the surrounding neutral gas, leading to lowering of \hi\ column density and detection rate. Further, \citet{curran2013a} predicted a cut-off AGN UV luminosity of $\rm 10^{23} \ W \ Hz^{-1}$, above which they claim that all the neutral gas in the AGN surroundings would be ionised and \hi\ 21-cm absorption cannot be detected. 

It should be noted, however, that the observed \hi\ 21-cm absorption strength depends critically on the fraction of the background radio source that is covered by the foreground neutral gas. As mentioned in Section~\ref{sec:absorption_associated_lowz}, the \hi\ detection rate is known to be higher in compact AGNs compared to extended sources. In addition, the AGN orientation with respect to the gas disk and the gas spin temperature are strong factors that affect \hi\ 21-cm absorption strength. \citet{aditya2016} and \citet{aditya2018b} conducted a survey using GMRT in a sample of 92 uniformly-selected flat-spectrum radio sources at $0<z<3.6$ (see Fig.~\ref{fig_hiabs_zevolution}). The radio sources in their sample are either compact or are core-dominated, indicating that the gas covering factor is nearly uniform across the sample. They find a significant dependence (at $\gtrsim3\sigma$ significance) of \hi\ 21-cm absorption strength on redshift and UV and 1.4\,GHz radio luminosities of the AGNs, with lower \hi\ 21-cm absorption strength at higher redshifts and higher luminosities. While a high UV luminosity of the AGN would ionise the neutral gas as discussed above, a high radio luminosity would raise the gas spin temperature; both these effects would lower the \hi\ 21-cm absorption strength and detection rate. Further, a redshift evolution in the gas properties, for example higher spin temperatures in systems at high redshifts as observed in intervening DLAs (Section~\ref{sec:absorption_intervening_lya}), could also cause a low absorption strength at high redshifts. Based on non-detection of associated \hi\ 21-cm absorption in a sample of 29 AGNs at $0.7<z<1$ with uGMRT, \citet{murthy2022} found that the strength of the \hi\ absorption towards extended radio sources is significantly ($\approx3\sigma$) weaker at $0.7<z<1$ than at $z<0.25$, which they attribute to a redshift evolution in the physical conditions of the \hi\ gas due to either a low \hi\ column density or a high \ts\ in high-$z$ AGN environments.

\citet{grasha2019} conducted a survey of \hi\ 21-cm absorption in a sample of 145 compact sources at $0.02<z<3.8$. While their results confirm a low detection fraction at high redshifts, they do not find a correlation between \hi\ absorption detection rate and 1.4\,GHz luminosity. Further, they do not find any detections in AGNs with UV luminosities higher than $\rm 10^{23} \ W \ Hz^{-1}$, supporting the hypothesis of \citet{curran2013a}. However, \citet{aditya2017} and \citet{aditya2021} have reported detections of associated \hi\ 21-cm absorption towards two compact AGNs at $z\approx1.2$ and $z\approx3.5$ respectively, which have UV luminosities higher than $\rm 10^{23} \ W \ Hz^{-1}$, demonstrating that neutral gas can indeed survive in the host galaxies of AGNs with high UV luminosities. Future surveys that can compare the \hi\ 21-cm absorption properties in statistical samples with similar luminosities at different redshifts, and in samples at similar redshifts but with different luminosities, will be able to break the above degeneracy and test for redshift evolution. 

Furthermore, the presence of dust as indicated by red colours and strong \mgii\ absorption in the optical spectra of radio quasars have been observed to lead to an increased detection rate of associated \hi\ absorption \citep{carilli1998a,ishwarachandra2003,dutta2020a}. The \hi\ 21-cm absorption in such systems is typically observed to be kinematically offset from the \mgii\ absorption in the optical spectra of the quasars. This could be due to the \hi\ gas in the vicinity of these AGNs having small-scale structures that give rise to different absorbing components, and the optical and radio sightlines probing different volumes of the gas \citep{dutta2020a}.

\subsubsection{Spatially-resolved studies of \hi\ gas in AGN:}
\label{sec:absorption_associated_highres}

The definitive way to trace the circumnuclear regions in radio galaxies is through high-spatial resolution (mas) observations using VLBI. At $z\lesssim0.1$, \hi\ 21-cm absorption line can be used to trace the structure and kinematics of the circumnuclear regions if the radio continuum emission is sufficiently bright and extended at mas-scales \citep[see][for compilations of \hi\ VLBI observations of AGNs]{araya2010,morganti2018}. Such observations have uncovered a variety of circumnuclear \hi\ structures including discs, tori, rings and bars \citep[e.g.][]{taylor1996,carilli1998b,cole1998,peck1999,mundell2003,morganti2011}. In addition, high-spatial resolution observations reveal feeding and feedback phenomena associated with the AGN traced through \hi\ gas inflows and outflows, particularly in young radio sources. These observations show that most of the fast ($\gtrsim1000$\,\kms) and massive (upto few tens \msunyr) \hi\ outflows are driven by radio jets from the AGN \citep[e.g.][]{oosterloo2000,morganti2004,morganti2013,schulz2018,schulz2021}. Furthermore, these outflows often trace a clumpy medium and different stages in the interaction between radio jets and ISM clouds. It has been more challenging to obtain direct evidence of \hi\ gas fuelling the central AGN due to difficulty in distinguishing between infalling and rotating gas. However, few high-spatial resolution observations of \hi\ absorption do show signatures of infalling gas clouds fuelling the AGN activity \citep[e.g.][]{araya2010,struve2010,srianand2015}.

\section{\hi\ 21-cm science with SKA and its Pre-cursors/Pathfinders}
\label{sec:ska}

The unmatched sensitivity and resolution of the SKA \citep{dewdney2009} will allow us to address a wide variety of open questions in the field of galaxy evolution using the \hi\ 21-cm line. In preparation for SKA1, a large number of surveys with SKA pathfinders, e.g. upgraded GMRT \citep[uGMRT;][]{gupta2017}, APERture Tile In Focus \citep[APERTIF;][]{oosterloo2009}, and pre-cursors, e.g. Australian SKA Pathfinder \citep[ASKAP;][]{hotan2021}, MeerKAT \citep{jonas2016}, are being planned/conducted. We discuss here the prospects for \hi\ 21-cm emission and absorption science with the above facilities.

\subsection{\hi\ emission science}
\label{sec:ska_emission} 

Due to the intrinsic faintness of the \hi\ 21-cm emission line, \hi\ mapping of galaxies has been feasible only up to $z\sim$0.2 with the existing radio telescopes, except for a few objects that were detected using long integration times. The COSMOS \hi\ Large Extragalactic Survey (CHILES) has acquired the highest redshift direct \hi\ detection to date at $z\sim$0.376 \citep{fernandez2016}. More recently, there has been tentative detection of \hi\ emission from a strongly lensed galaxy at $z\sim$0.407 \citep{blecher19}. SKA pre-cursors are extending the scope of \hi\ 21-cm emission studies of galaxies by improving the observational capabilities in radio astronomy such as sensitivity, survey speed and resolution. We discuss here some of the major ongoing/upcoming \hi\ 21-cm emission surveys with ASKAP and MeerKAT that include: (i) wide-area shallow surveys for cosmology, (ii) medium-area deep surveys to study individual galaxies and their environmental dependence, and (iii) medium to narrow-area deep surveys to study the redshift evolution of galaxies.

The Widefield ASKAP L-band Legacy All-sky Blind surveY \citep[WALLABY;][]{koribalski20}, the wide-area shallow survey being carried out using ASKAP, proposes to observe $\sim75$\% of the sky ($-90^{\circ} < \delta < 30^{\circ}$) up to $z \sim 0.26$. It is estimated that around 500000 galaxies will be detected over the entire survey area (assuming 30$''$ angular resolution), with $\sim$1000 of them being spatially well-resolved (with $>$10 beams). This survey will give a complete 3D sampling of the southern sky, which can be used to refine cosmological parameters using the spatial and redshift distribution of low-mass gas-rich galaxies. WALLABY Early Science results \citep{reynolds19, leewaddell19, elagali19, kleiner19, for19} include the discovery of new dwarf galaxies and \hi\ debris in galaxy groups.

The \hi\ component of the MeerKAT International GHz Tiered Extragalactic Exploration (MIGHTEE) survey is a medium-deep, medium-wide survey that covers a 20\,deg$^2$ area out to $z \sim 0.6$ \citep{jarvis16, maddox21}. The combination of MeerKAT data and existing extensive multiwavelength data would allow studies of the (i) HIMF as a function of redshift, (ii) \hi\ kinematics including BTFr and dark matter distribution of dwarfs, (iii) \hi\ gas properties as a function of environment, (iv) redshift evolution of the neutral gas content over the past 5 billion years, and (iv) spectral stacking to detect \hi\ below the detection limit.

The MeerKAT Fornax Survey \citep[MFS;][]{serra16} is a medium-deep survey to observe a region of $\sim$12\,deg$^2$, reaching a projected distance of 1.5\,Mpc from the centre of the Fornax cluster. This survey will cover a wide range of environments from cluster centre to outskirts, which will allow studies of the HIMF down to $10^{5}$\,M$_{\odot}$ as a function of the environment. In addition, the \hi\ morphologies of resolved galaxies (at resolution $\sim 1$\,kpc and \nhi\ sensitivity of few times $10^{19}$\,\cms) in the Fornax cluster can be used to study the environmental processes as a function of cluster-centric radius.

The MeerKAT Observations of Nearby Galactic Objects - Observing Southern Emitters \citep[MHONGOOSE;][]{deblok16} survey is a medium-deep pointed MeerKAT \hi\ survey of 30 nearby galaxies (55\,h per field) to study the low-column density environment down to a $3\sigma$ \nhi\ limit of below $10^{18}$\,\cms, which is a few hundred times fainter than the typical \hi\ disks in galaxies. This survey is expected to directly detect the effects of cold accretion from the IGM and the links with the cosmic web. This is the first survey to combine high \nhi\ sensitivity and high spatial resolution over a large field of view, and will therefore probe the connection between galaxies and the IGM in the nearby Universe at an unprecedented level. 

The Deep Investigation of Neutral Gas Origins \citep[DINGO;][]{meyer09,duffy12} survey is a deep medium-wide survey that aims to observe $\sim$180\,deg$^2$ of the sky up to $z \sim 0.4$ with ASKAP. The survey targets fields covered by the Galaxy And Mass Assembly survey \citep{driver2011}, and therefore has rich ancillary data including galaxy redshifts, stellar and star formation properties, and environmental catalogues. The full DINGO survey aims to detect all galaxies with M$_{\rm HI} > 10^9$ M$_{\odot}$ within the survey area up to $z \lesssim 0.1$ in order to sample below the knee of the HIMF. Through stacking, DINGO aims to detect \hi\ in different halo mass (M$_{\rm h} = 10^{12} - 10^{14}$ M$_{\odot}$) and redshift ($0.25 < z < 0.4$) bins. In total, DINGO expects to detect $>10^4$ galaxies up to $z<0.4$. 

The Looking at the Distant Universe with MeerKAT Array  \citep[LADUMA;][]{holwerda12, blyth16} survey being carried out using MeerKAT will be the deepest \hi\ 21-cm emission survey. Uisng 3424\,h of observations in a single pointing in the Extended Chandra Deep Field South, this survey will probe the \hi\ gas in emission for the first time up to $z\sim1.4$ through direct and stacked detections. LADUMA will study the \hi\ mass function and the connection between \hi\ gas and galaxy properties as a function of redshift and environment, and the redshift evolution of the BTFr and the cosmic \hi\ density.

Looking further ahead, the SKA will provide breakthrough observations with its unprecedented sensitivity, sky coverage and spatial resolution. The full SKA will allow mapping of individual galaxies at 1'' resolution out to $z\sim1$, an order of magnitude deeper than the current interferometers, and reach the same angular resolutions that are now achieved for the Local Group out to $\sim$10 Mpc. In addition, the SKA will provide enough survey speed to detect \hi\ emission from around a billion galaxies over 3/4$^{th}$ of the sky out to $z\sim2$ \citep{yahya15}. The SKA will provide the possibility to study tidal interactions and minor mergers in unprecedentedly large galaxy samples, hence allowing us to investigate the accretion and removal rates from interaction, and their dependence on environment and time. Further, the SKA will allow the study of kinematics of \hi\ halos, which gives essential information about the exchange of material and angular momentum between galactic disks and the IGM \citep{marinacci10}. With the SKA, it will be possible to study the \hi\ mass function up to higher redshifts ($z\sim1$). The slope, scatter and zero-point of the BTFr can be studied as a function of redshift, cosmic environment and global galaxy properties with the SKA, since it provides adequate angular resolution and column density sensitivity to image the \hi\ kinematics at higher redshifts, as well as survey volumes that encompass all cosmic environments.

\subsection{\hi\ absorption science}
\label{sec:ska_absorption}

Despite extensive efforts by the community as detailed in Section~\ref{sec:absorption}, the relatively small number of \hi\ 21-cm absorbers detected to date ($\approx60$ intervening absorbers; $\approx150$ associated absorbers, only 12 at $z>1$) hinders our ability to utilise these absorbers to trace the evolution of neutral gas and its interplay with galaxies across cosmic time. The main technical limitations affecting \hi\ 21-cm absorption searches so far have been small bandwidths and restricted frequency ranges due to the presence of strong radio-frequency interference (RFI). It has thus been challenging to conduct blind absorption line searches in statistical samples \citep[see however][]{darling2011,grasha2020}, leading to biases due to small sample sizes and different optical/UV pre-selection techniques that complicate the interpretation of results. The instantaneous large bandwidths, broad frequency coverage in low-RFI environments and high survey speed of the new telescopes are enabling large, blind and wide-area surveys of \hi\ 21-cm absorption, e.g. Search for \hi\ absorption with APERTIF \citep[SHARP;][]{maccagni2017}, First Large Absorption Survey in \hi\ \citep[FLASH;][]{allison2021} and MeerKAT Absorption Line Survey \citep[MALS;][]{gupta2016}. SHARP will search for \hi\ 21-cm absorption up to $z\approx0.26$ in the northern hemisphere using the new APERTIF system on WSRT. In the southern hemisphere, FLASH will search for \hi\ 21-cm absorption over the intermediate redshift range, $0.5<z<1.0$, using ASKAP, while MALS will search for absorption over $0<z<1.5$ using MeerKAT.

By searching for absorption along tens of thousands of sightlines towards radio sources down to $\approx15$\,mJy over $\approx1000-30000$\,deg$^2$, the above surveys are expected to produce several hundreds of intervening absorbers and thousands of associated absorbers out to $z\approx1.5$. These surveys will not just expand the sample sizes of \hi\ 21-cm absorbers by over an order of magnitude, being blind surveys they will enable us to trace the cosmic evolution of cold neutral gas content in galaxies in a systematic and unbiased way using homogeneous flux-limited samples. Comparison of \hi\ absorbing gas properties across different classes of normal and active galaxies with high statistical confidence will also become possible with such large samples. Early science results are already beginning to demonstrate the potential of these large surveys, e.g. searches towards dust-obscured, reddened, lensed and radio-selected sources \citep{glowacki2019,combes2021,mahony2022}, new constraints on the column density distribution, number density and covering factor of \hi\ absorbers from blind samples \citep{allison2020,sadler2020,gupta2021a}.

In addition, the increased sensitivity and almost continuous frequency coverage over 125-1460\,MHz offered by the uGMRT is allowing us to conduct blind \hi\ 21-cm absorption surveys up to higher redshifts and in redshift ranges that were previously inaccessible to radio interferometers in relatively RFI-free environments. For example, uGMRT has enabled new detections of intervening \hi\ 21-cm absorbers in $z\approx2$ DLAs \citep{kanekar2014b} and associated \hi\ 21-cm absorbers in the intermediate redshift range $0.7<z<1.0$ \citep{aditya2019,murthy2022}, at $z\approx1$ \citep{dutta2020a,chowdhury2020a}, at $z\approx2$ \citep{gupta2021b} and at $z\approx3.5$, the highest redshift at which \hi\ 21-cm absorption has been detected to date \citep{aditya2021}. Based on blind, dust-unbiased surveys with uGMRT at $z<0.4$ and at $2<z<5$, the number of \hi\ 21-cm absorbers per unit redshift is estimated to be $<0.14$ at $z\approx0.18$ and $<0.048$ at $z\approx2.5$ for $5\sigma$ \nhi\ threshold of $5\times10^{19}$\,\cms\ and \ts\ = 100\,K \citep{gupta2021a,gupta2021b}. These surveys also constrain the covering factor of CNM gas to be 0.022 at impact parameters between 50 and 150\,kpc from $z<0.4$ galaxies and $\le0.2$ in $2<z<3$ DLAs. Further, from an unbiased survey of \hi\ 21-cm absorption over $0.7<z<1.5$ towards radio sources in the DEEP2 fields, \citet{chowdhury2020a} detected two of the highest optical depth \hi\ 21-cm absorbers (\nhi\ $\gtrsim10^{22}$\,\cms) at $z\approx1.2$ that are associated with reddened AGNs with low rest-frame radio and UV luminosities.

Building upon the knowledge gained through the above studies, ongoing and future \hi\ 21-cm absorption surveys with the pathfinders, pre-cursors and eventually SKA, will put more stringent constraints on cold \hi\ gas properties in galaxies and their redshift evolution by covering larger redshift pathlengths and will also open up new discovery space. Proposed \hi\ 21-cm absorption surveys with SKA1-MID (350-1050\,MHz) are expected to reach spectral line sensitivities of $<0.1$\,mJy and detect several thousand intervening absorbers and a few hundred outflows over $z\approx$ over 10000\,deg$^2$ up to $z\approx3$, while surveys with SKA1-LOW (50-350\,MHz) are expected to reach sensitivities of $<0.5$\,mJy and discover new galaxies and quasars through \hi\ 21-cm absorption over the as of yet poorly explored redshift range of $3<z<8$ \citep{morganti2015a}. Subsequently, SKA2 will facilitate measurements of mass, size and kinematics of normal galaxies and detailed studies of cold gas flows in radio galaxies over almost the full range of cosmic time \citep{kanekar2004}. 
As a final point, it is noted that the SKA will enable us to test the importance of the role of magnetic fields in galaxy formation and evolution through measurement of Zeeman splitting and rotation measure synthesis in MW and external galaxies \citep{robishaw2015,heald2020}. For example, observations of Faraday rotation measure and Zeeman splitting in the \hi\ 21-cm absorption line arising from DLAs and \mgii\ systems could be used to constrain the evolution of magnetic fields in galaxies across cosmic time \citep{farnes2017}.

\section{Conclusion}
\label{sec:conclusions}

In this review we have discussed different \hi\ 21-cm emission and absorption studies of the atomic hydrogen gas in galaxies, which fuels and regulates star formation in galaxies along with being a key input to understanding how various physical processes govern galaxy formation and evolution. Further ahead, the SKA will revolutionize our knowledge of the role of \hi\ gas in galaxy evolution. Studying the neutral gas distribution and kinematics of galaxies in statistical samples, in different environments and as a function of cosmic time is vital to develop a holistic understanding of galaxy evolution. Indeed, this is one of the key science drivers for the development of the SKA. The combination of unprecedented sensitivity, resolution and wide receiver bands of the SKA will permit us to image the \hi\ gas in emission in galaxies in the Local Volume at resolutions of $<100$\,pc and in a substantial number of high-$z$ galaxies for the first time. Complementary to sensitive \hi\ emission observations, blind, wide-area \hi\ absorption line surveys will uncover several thousands of intervening and associated absorbers that will trace the \hi\ gas in galaxies up to the earliest cosmic epochs.

Finally, \hi\ 21-cm surveys with SKA pre-cursors and eventually SKA will provide exciting avenues for multi-wavelength follow-up studies to combine emission and absorption line observations and connect multiphase gas in and around galaxies \citep[e.g.][]{morganti2015b,neeleman2017,kanekar2018b,combes2019,peroux2019,weng2021}. There will be ample scope for synergy with recent and upcoming telescopes in other wavelengths that will trace different gas phases e.g. cold molecular gas using (sub-)mm observations with the Atacama Large Millimeter/submillimeter Array and the Northern Extended Millimeter Array, warm ionized gas using optical/IR observations with the Extremely Large Telescope, the James Webb Space Telescope and the Thirty Meter Telescope, and hot ionized gas using X-ray observations with eROSITA and Athena. Therefore, in the coming few decades we will be able to address some of the key open questions in galaxy evolution: what is the role played by \hi\ gas in the evolution of galaxies, how does the connection between \hi\ gas, other gas phases and galaxy properties vary as a function of galaxy environment and cosmic time, how do gas flows drive the co-evolution of AGN and galaxies over cosmic time.

\section*{Acknowledgements}

We thank the anonymous reviewer for useful comments that helped improve the quality of this review article.
RD gratefully acknowledges support from the European Research Council (ERC) under the European Union’s Horizon 2020 research and innovation programme (grant agreement No 757535). SK is supported by the South African Research Chairs Initiative of the Department of Science and Technology and National Research Foundation.


%
\def\aap{A\&A}%
\def\aapr{A\&A~Rev.}%
\def\aaps{A\&AS}%
\def\aj{AJ}%
\def\actaa{Acta Astron.}%
\def\araa{ARA\&A}%
\def\apj{ApJ}%
\def\apjl{ApJ}%
\def\apjs{ApJS}%
\def\apspr{Astrophys.~Space~Phys.~Res.}%
\def\ao{Appl.~Opt.}%
\def\aplett{Astrophys.~Lett.}%
\def\apss{Ap\&SS}%
\def\azh{AZh}%
\def\bain{Bull.~Astron.~Inst.~Netherlands}%
\def\baas{BAAS}%
\def\bac{Bull. astr. Inst. Czechosl.}%
\def\caa{Chinese Astron. Astrophys.}%
\def\cjaa{Chinese J. Astron. Astrophys.}%
\def\fcp{Fund.~Cosmic~Phys.}%
\def\gafd{Geophys.\ Astrophys.\ Fluid Dyn.}
\def\gca{Geochim.~Cosmochim.~Acta}%
\def\grl{Geophys.~Res.~Lett.}%
\def\iaucirc{IAU~Circ.}%
\def\icarus{Icarus}%
\def\jcap{J. Cosmology Astropart. Phys.}%
\def\jcp{J.~Chem.~Phys.}%
\def\jfm{JFM}
\def\jgr{J.~Geophys.~Res.}%
\def\jqsrt{J.~Quant.~Spec.~Radiat.~Transf.}%
\def\jrasc{JRASC}%
\def\mnras{MNRAS}%
\def\memras{MmRAS}%
\def\memsai{Mem.~Soc.~Astron.~Italiana}%
\def\na{New A}%
\def\nar{New A Rev.}%
\def\nat{Nature}%
\def\nphysa{Nucl.~Phys.~A}%
\def\pasa{PASA}%
\def\pasj{PASJ}%
\def\pasp{PASP}%
\def\physrep{Phys.~Rep.}%
\def\physscr{Phys.~Scr}%
\def\planss{Planet.~Space~Sci.}%
\def\pra{Phys.~Rev.~A}%
\def\prb{Phys.~Rev.~B}%
\def\prc{Phys.~Rev.~C}%
\def\prd{Phys.~Rev.~D}%
\def\pre{Phys.~Rev.~E}%
\def\prl{Phys.~Rev.~Lett.}%
\def\procspie{Proc.~SPIE}%
\def\qjras{QJRAS}%
\def\rmxaa{Rev. Mexicana Astron. Astrofis.}%
\def\sgg{Stud.\ Geoph.\ et\ Geod.}
\def\skytel{S\&T}%
\def\solphys{Sol.~Phys.}%
\def\sovast{Soviet~Ast.}%
\def\ssr{Space~Sci.~Rev.}%
\def\zap{ZAp}%
\def\memsai{Memorie della Societa Astronomica Italiana}
\def\araa{Ann. Rev. of Astron. and Astrophys.}
\def\zhetp{JETP}
\def\jetp{Sov.\ Phys.\ JETP}
\let\astap=\aap
\let\apjlett=\apjl
\let\apjsupp=\apjs
\let\applopt=\ao

\bibliography{mybib}

\end{document}